\documentclass[12pt]{article}

\usepackage{amssymb}
\usepackage{amsmath}

\usepackage{relsize}
\usepackage{booktabs}
\usepackage{pgfplots}
\usepackage{placeins}
\usepackage{indentfirst}
\usepackage{multirow}
\usepackage[english]{babel}
\usepackage [autostyle, english = american]{csquotes}
\usepackage[margin=1.25in]{geometry}
\usepackage{setspace}
\usepackage{graphicx} 
\usepackage{relsize}
\usepackage{tikz}

\usetikzlibrary{calc,angles,positioning,intersections,quotes,decorations.markings}

\usepackage{pgfplots}
\usepackage{mathtools}

\usepackage{hyperref}
\hypersetup{
    colorlinks=true,
    linkcolor=blue,
    citecolor=blue,
    filecolor=magenta,      
    urlcolor=cyan,
}
\usepackage{apacite}

\newcommand{\verteq}{\rotatebox{90}{$\,=$}}
\newcommand{\equalto}[2]{\underset{\scriptstyle\overset{\mkern4mu\verteq}{#2}}{#1}}
\title{Rational Inattention and Perceptual Distance}
\author{David Walker-Jones
\\ University of Toronto
\\david.walker.jones@mail.utoronto.ca
\\Keywords: rational inattention, Shannon Entropy, perceptual distance.
}

\begin{document}
\maketitle
\setlength{\parindent}{1cm}

\subsection*{Abstract}

\par
\begin{spacing}{1}
This paper uses an axiomatic foundation to create a new measure for the cost of learning that allows for multiple perceptual distances in a single choice environment so that some events can be harder to differentiate between than others. The new measure maintains the tractability of Shannon's classic measure but produces richer choice predictions and identifies a new form of informational bias significant for welfare and counterfactual analysis.\footnote{Special thanks to Rahul Deb for all of the support. I would also like to thank Yoram Halevy, Marcin Peski, Carolyn Pitchik, and Colin Stewart, for their helpful advice.}
\end{spacing}

\par

\section{Introduction}

In many choice environments it is costly for agents to obtain information about the options that they face. Understanding how agents learn in such environments is crucial because partially informed choices have serious implications for revealed preference analysis, which makes welfare and counterfacutal analysis more difficult.

The standard technique for quantifying the cost of learning in models of rational inattention (RI) is Shannon Entropy \cite{sim03}. Shannon Entropy has an axiomatic foundation, is grounded in the optimal coding of information, and provides a tractable and flexible framework with which to study agent behavior \cite{sha48}.

While Shannon Entropy has proven to be a valuable tool, it does have limitations in economic environments as they are not what it is designed for. It is natural to think, for instance, that it should be more difficult to differentiate between outcomes that are more similar. Differentiating between two types of black tea should be more difficult than differentiating between water and coffee\hypertarget{fo2}{.} Shannon Entropy, however, does not allow for different outcomes to be more or less similar than each other. Without a mechanism to allow for what is referred to in the literature as `perceptual distance,'\footnote{If two outcomes are more similar it is said that they have less perceptual distance between them.} the choice behavior predicted by Shannon Entropy can differ from observed behavior, as is demonstrated by \hyperlink{sec2.1}{Example 1} in \hyperlink{sec2.1}{Section 2.1}, which can limit the effectiveness of Shannon Entropy in empirical settings.

This paper proposes five axioms that focus on the cost of asking simple questions, questions that can be represented by partitions of the state space. The axioms are used to create a new measure for the cost of learning that we call Multisource Shannon Entropy (MSSE). MSSE features perceptual distance, maintains the desired tractability and flexibility of Shannon's classic measure when incorporated into a model of RI, and predicts behavioral patterns that have been identified as problematic for Shannon Entropy.

MSSE also identifies a previously undiscovered informational bias in random utility (RU) models that should be considered a natural consequence of different perceptual distances in the same choice environment, as is demonstrated by \hyperlink{sec2.2}{Example 2} in \hyperlink{sec2.2}{Section 2.2}. While other papers study measures of information that feature perceptual distance \cite<e.g.,>{hebwoo17}, this paper is the first to identify informational biases in RU models generated by the presence of different perceptual distances in the same choice environment. Unlike the informational bias identified with Shannon Entropy, this type of informational bias cannot be identified in the unconditional choice probabilities of the agent, and thus presents a new challenge for welfare and counterfactual analysis.

\subsection{Literature Review}

Shannon Entropy has been used in several contexts to demonstrate informational biases in RU models. \citeA{matmck15} use Shannon Entropy in a model of RI to demonstrate the potential for informational biases in multinomial logit, while \citeA{steea17} use Shannon Entropy in a model of RI to demonstrate the potential for a similar bias in dynamic logit. These results are significant for those who wish to fit RU models because, while observational data may coincide with the assumptions of a fitted RU model, informational biases can potentially invalidate counterfactual and welfare analysis, two common goals of such a fitting. 


The Shannon RI model has also led to a number of predictive successes. \citeA{achwee16} show that using Shannon Entropy to model firms as rationally inattentive results in a better fitting of labor market dynamics after the great depression. \citeA{dasmon17} show that using Shannon Entropy to model importers as rationally inattentive results in novel predictions that are supported by trade data. \citeA{ambea18} experimentally verify predictions of Shannon Entropy in environments where agents are rationally inattentive to the consequences of participating in different transactions.

Perhaps as a response to the success Shannon Entropy has enjoyed, several recent papers have noted that Shannon Entropy may be a poor measure of the cost of acquiring information in some environments \cite{capea17, moryan16} because it lacks what is called ``perceptual distance'' \cite[p.~39]{capea17}. As was alluded to previously, these papers argue that (i) more similar outcomes (outcomes that have less perceptual distance between them) should be more difficult to differentiate between, and (ii) when this property is missing, predicted behavior can differ signficantly from the type of behavior that it would seem natural to expect \cite{moryan16}.

To better understand the relationship between the cost of learning and agent behavior, a number of papers have studied axiomatic models of rational inattention. Different papers, however, choose to focus their axioms on different aspects of the choice environment. \citeA{capea17}, for instance, develop axioms that focus on the choice behavior of an agent after they expend effort to learn about the state of the world. In contrast, \citeA{deo14} and \citeA{deoea17} develop axioms that focus on an agent's preferences over choice menus before they expend effort to learn about the state of the world. Broadly, these papers aim to understand what implications rational agent behavior has for the form of information cost functions. 

\citeA{ell18} features axioms that focus on choice behavior and studies the implications for information cost functions, but further assumes that the agent learns by picking a partition of the state space. While MSSE uses the cost of learning the realized event of partitions as a primitive, the model studied in this paper does not constrain agents so that they must learn using partitions of the state space, and it can be shown that it is never optimal for the agent to choose an information strategy that is equivalent to a partition of the state space \cite{wal19b}.\footnote{This is true whenever the agent does some learning, and they have a positive probability of a posterior that is different than their prior.}

Closer in nature to the work done in this paper, \citeA{pomea19} develop axioms that focus directly on the costs of information. Axioms that focus on costs of information are interesting because intuitive properties for costs of information can lead to unintuitive agent behavior that is compelling given real-world observations \cite{gigtod99}, but is often mistaken for irrational when axioms that appear rational are imposed on behavior. MSSE, for instance, predicts `non-compensatory' behavior, whereby changing an option so that it is more valuable to the agent can result in a lower chance of it being selected, as is discussed by \cite{wal19b}. This type of behavior raises important questions for welfare and counterfactual analysis, making effective policy design more challenging. 

Unlike the work of \citeA{pomea19}, which features axioms that are concerned with probabilistic experiments that can result in different outcomes in the same state of the world, this paper's axioms are concerned with deterministic experiments (questions) that always result in the same outcome in a given state of the world, and contradict the form of constant marginal cost assumed in their paper.

\subsection{Organization of Paper}
The remainder of the paper is organized as follows: \hyperlink{sec2}{Section 2} introduces Shannon Entropy, discusses models of RI, and provides motivating examples. \hyperlink{sec3}{Section 3} proposes five new axioms, and uses them to develop a more flexible cost of acquiring information, MSSE, which features perceptual distance. \hyperlink{sec4}{Section 4} uses MSSE as a benchmark with which to price inattentive information strategies in a model of RI, and discusses the resultant agent behavior. \hyperlink{sec5}{Section 5} discusses the relationship between RU models and the agent behavior found in \hyperlink{sec4}{Section 4}, and revisits the motivating examples from \hyperlink{sec2.1}{Section 2.1} and \hyperlink{sec2.2}{Section 2.2}. \hyperlink{sec6}{Section 6} concludes.

\hypertarget{sec2}{ }
\section{Rational Inattention and Shannon Entropy}

What follows is intended to introduce the Shannon Entropy model of rational inattention to those that are not familiar with it. If you are familiar with said model, you can skip to \hyperlink{sec2.1}{Section 2.1}.

In the rational inattention (RI) literature, learning by the agent is typically modelled as the choice of a signal structure. The agent chooses the probability of receiving different signals in different states of the world. Receiving a signal updates the agent's belief about the state of the world, giving them a more informed posterior belief. More informative signal structures are more costly for the agent, but allow them to make a more informed decision about which option to select.

Suppose that the uncertainty faced by the agent is described by a measurable space $(\Omega, \, \mathcal{F})$, where $\Omega$ is a finite set of possible \textbf{states of the world} (the state space), and $\mathcal{F}$ is the set of \textbf{events} generated by $\Omega$ (the power set of $\Omega$). We call $\mu:\mathcal{F}\rightarrow [0,\,1]$, which assigns probabilities to events, the \textbf{prior} distribution of the agent.

Suppose that an agent who has stopped learning must make a selection from a set of \textbf{options}, denoted $\mathcal{N}=\{1,\,\dots,\,N\}$. Each option, $n\in\mathcal{N}$, in each state of the world, $\omega\in\Omega$, has a (finite) \textbf{value} to the agent $\textbf{v}_n(\omega)$.

The agent's problem is to maximize the expected value of the selected option less the cost of learning. They do this by choosing an \textbf{information strategy} $F(s, \omega)\in \Delta(\mathbb{R} \times \Omega)$, which is a joint distribution between $s$, the observed \textbf{signal}, and the states of the world.\footnote{The decision to allow $s$ to be any real number is rather arbitrary. This is a much richer signal space than is required in practice. We show later that an optimal strategy only results in one of at most $N$ different signals $s$ being observed.} The only restriction on the information strategy is that the marginal, $F(\omega):\mathcal{F}\rightarrow \mathbb{R}_+$, must equal the prior $\mu$. Alternatively, an agent can select a probability measure $F(s|\omega) : \mathbb{R} \rightarrow \mathbb{R}_+$ for each $\omega\in \Omega$, which, combined with $\mu$, determine both $F(s,\,\omega)$ and the posterior $F(\omega|s)$. It is a property of the cost function for information derived in this paper, as is true with Shannon Entropy, that if $F(s, \omega)$ is optimal, then the agent is done learning after a single signal $s$. After the signal is realized, the agent simply picks the action with the highest expected value:
$$
a(s|F)=\text{arg}\max_{n\in \mathcal{N}} \mathbb{E}_{F(\omega|s)}[\textbf{v}_n(\omega)].$$
\noindent Ignoring the cost of learning momentarily, the value to the agent of receiving a signal $s$, which induces posterior $F(\omega|s)$, is then:
$$V(s|F) = \max_{n \in \mathcal{N}}  \mathbb{E}_{F(\omega|s)}[\textbf{v}_n(\omega)].
$$

Let the expected cost of a particular information strategy, given the agent's prior, be denoted $\textbf{C}(F(s,\,\omega),\,\mu)$. We describe the form of the cost functions studied in this paper in \hyperlink{sec4}{Section 4}. The agent's problem can thus be written:
$$
\max_{F\in \Delta(\mathbb{R} \times \Omega)} \sum\limits_{\omega\in\Omega}\int\limits_s V(s|F)F(ds|\omega)\mu(\omega) - \textbf{C}(F(s,\,\omega),\,\mu),
$$ 
$$
\text{such that } \forall \omega \in \Omega:\, \int\limits_s F(ds,\,\omega) = \mu(\omega).
$$

The choice behavior the agent exhibits depends on the cost function for information. Shannon Entropy is a measure of uncertainty with an axiomatic foundation that can be used to assign costs to information. If we are given a partition of the possible states of the world $\mathcal{P}=\{A_1,\,\dots,\,A_m\}$\hypertarget{fo3}{,} and probability measure $\mu$ over these events, the uncertainty about which event has occurred, as measured by \textbf{Shannon Entropy}, is defined:\footnote{This measure is only unique up to a positive multiplier.}

\hypertarget{1}{}
\begin{equation}
\mathcal{H} (\mathcal{P},\,\mu) = - \sum\limits_{i=1}^m \mu(A_i) \log(\mu(A_i)).
\end{equation}

\noindent The convention used here is to set $0\log(0)=0$. 

If an agent has prior $\mu$ about the state of the world, and their beliefs are updated to the posterior $\mu(\cdot|s)$ after they receive a signal $s$, then there is a change in the uncertainty as measured by Shannon Entropy. In the Shannon model of RI, the cost of an information strategy $F(s,\omega)$ is measured as the expected reduction in total uncertainty as measured by Shannon Entropy:
$$\mathbb{E}\Big[\mathcal{H} (\mathcal{P},\,\mu) - \mathcal{H} (\mathcal{P},\,\mu(\cdot|s))\Big],
$$
\noindent where $\mathcal{P}=\{\{\omega_1\},\,\{\omega_2\},\,\dots,\,\{\omega_n\}\}$. Bayes rule, and the nature of Shannon Entropy, guarantee that every potential information strategy has a weakly positive cost.

\hypertarget{sec2.1}{ }

\subsection{Example 1: Perceptual Distance and Problems with Predictions}

\citeA[p.~19]{capea17} show that Shannon Entropy results in choice behavior that satisfies ``invariance under compression.'' That is, when Shannon Entropy is used to measure information, if there are two states of the world, $\omega_1$ and $\omega_2$, across which payoffs are identical for each option ($\textbf{v}_n(\omega_1) = \textbf{v}_n(\omega_2)\,\,\forall n\in\mathcal{N}$), then the chance of each option being selected is the same in $\omega_1$ and $\omega_2$. The invariance under compression that is predicted by Shannon Entropy is, unfortunately, not found in many settings, as is shown by the work of \citeA{deanel17}. The intuition for why invariance under compression may not be present in every choice environment is demonstrated by the following example.

Consider an environment where an agent is faced with a screen that shows 100 balls, each of which is either red or blue. The agent is offered a prize that they may either accept (option 1), or reject to get a payoff of zero (option 2). The agent is told that if the majority of the balls on the screen are blue then the prize is $y\in\mathbb{R}_{++}$, and if the majority of the balls on the screen are red then the prize is $-y$. Suppose further that the agent is also told that there is a 1/4 chance of four different states of the world in which there are either 40, 49, 51, or 60 red balls, as is described in \hyperlink{ta1}{Table 1}.

\begin{table}[]
\hypertarget{ta1}{ }
\centering
\resizebox{\columnwidth}{!}{%
\label{my-label}
\begin{tabular}{|c|c|c|c|c|}
\hline
\multicolumn{5}{|c|}{Table 1: \hyperlink{sec1.1}{Example 1}}   \\ \hline
    State:        & $\omega_1$ & $\omega_2$ & $\omega_3$ & $\omega_4$      \\ \hline
   Balls in State:& 60 Blue \& 40 Red & 51 Blue \& 49 Red & 49 Blue \& 51 Red & 40 Blue \& 60 Red     \\ \hline
Probability of State: & 1/4 & 1/4 & 1/4 &  1/4   \\ \hline
Value of selecting option $1$:   & $y$ & $y$ & -$y$ &  -$y$    \\ \hline
Value of selecting option $2$:   & 0 & 0 & 0 & 0    \\ \hline
\end{tabular}
}
\end{table}

The Shannon RI model, which imposes invariance under compression, predicts that the agent has the same chance of selecting option $1$ when there are 40 red balls as when there are 49 red balls, and that the agent has the same chance of selecting option $1$ when there are 60 red balls as when there are 51 red balls. This predicted behavior is not intuitive because it should be easier for the agent to differentiate between the states that are more different (40 versus 60 red balls) than the states that are more similar (49 versus 51 red balls). One should instead expect that the chance that option 1 is selected is decreasing in the number of red balls, as is demonstrated by the experiments of \citeA{deanel17}, because it should be easier to determine which color of ball constitutes the majority the more of that color ball there are.

Why does Shannon Entropy impose this type of behavior? In short, Shannon Entropy results in invariance under compression because of Shannon's third axiom \cite{sha48}. In the context of \hyperlink{sec2.1}{Example 1}, let $\mathcal{P}=\{\{\omega_1\},\,\{\omega_2\},\,\{\omega_3\},\,\{\omega_4\}\}$, and $\tilde{\mathcal{P}}=\{\{\omega_1\cup\omega_2\},\,\{\omega_3\cup\omega_4\}\}$, be two partitions of the state space. Shannon's third axiom requires that total uncertainty about the state of the world, which is the uncertainty about which event in $\mathcal{P}$ has occurred, be equal to the uncertainty about which event in $\tilde{\mathcal{P}}$ has occurred, plus the expected amount of uncertainty that remains about which event in ${\mathcal{P}}$ has occurred after we have learned which event in $\tilde{\mathcal{P}}$ has occurred. This equality means that the reduction in uncertainty caused by a signal is equal to the reduction in uncertainty about which event in $\tilde{\mathcal{P}}$ has occurred, plus the expected reduction in uncertainty about which event in ${\mathcal{P}}$ has occurred given which event in $\tilde{\mathcal{P}}$ has occurred.

The agent is only concerned with which event in $\tilde{\mathcal{P}}$ has occurred, as this fully determines payoffs. Given which event in $\tilde{\mathcal{P}}$ has occurred, the agent does not care which event in ${\mathcal{P}}$ has occurred. If agent behavior is different in $\omega_1$ compared to $\omega_2$, or $\omega_3$ compared to $\omega_4$, so that their behavior does not satisfy invariance under compression, then the agent is, to an extent, differentiating between these states, and paying for information that does not benefit them, and their information strategy is thus not optimal.

Whil\hypertarget{fo4}{e} other information cost functions do not require that choice behavior satisfies invariance under compression \cite{capea17, moryan16}, they lack the tractability and flexibility of Shannon Entropy,\footnote{Shannon Entropy has a number of mathematical properties that make it easy to use for predicting behavior in a wide range of environments.} which limits the potential for their application. This has led to the following open question: ``what workable alternative models allow for the complex behavioral patterns identified in practice?'' \cite[p.~2]{capea17}, a question that this paper attempts to answer.

\hypertarget{sec2.2}{ }

\subsection{Example 2: Perceptual Distance and Biases in Fitting}

If different perceptual distances are present in the same choice environment, a RU model may be susceptible to a form of informational bias that has not previously been identified, as demonstrated by the following example. This is significant for those who wish to conduct welfare or counterfactual analysis because there are many economically significant examples where, for instance, one option is easier to learn about, as in \hyperlink{sec2.2}{Example 2}.

Consider a choice environment where an agent has two options: option 1 and option 2, which can each be of high value $H$, or low value $L<H$, as is described in \hyperlink{ta2}{Table 2}. Assume, contrary to what is possible with Shannon Entropy\hypertarget{fo5}{,} that learning the value of option $1$ is less costly than learning the value of option $2$.\footnote{With Shannon Entropy it is not possible for the cost of learning the value of option 1 to differ from the cost of learning the value of option 2. Each option realizes each of its two values with equal probabilities, and with Shannon Entropy it is not possible to have different perceptual distances in the same choice environment.} For example, perhaps the agent is interested in investing in one of two businesses that are \textit{a priori} identical except for the fact that one is local and easier to learn about, while the other is foreign and harder to learn about. It is not difficult to come up with other similar examples.

Because payoffs are symmetric, any knowledge about the value of option $1$ has the same value to the agent as the same knowledge about option $2$. Further, the cost of said information about option $1$ is lower. As such, while the marginal benefit of information about option $1$ or option $2$ is the same, the marginal cost of information about option $1$ is lower. We should thus expect research of a rational agent to be more attentive to option $1$. If the agent was deciding between investing in two businesses that are \textit{a priori} identical, except one is local and easier to learn about, while the other is foreign and harder to learn about, then we should expect the agent to be more attentive to the local business.

\begin{table}[]
\hypertarget{ta2}{ }
\centering
\label{my-label}
\begin{tabular}{|c|c|c|c|c|}
\hline
\multicolumn{5}{|c|}{Table 2: \hyperlink{sec1.2}{Example 2}}   \\ \hline
    State:        & $\omega_1$ & $\omega_2$ & $\omega_3$ & $\omega_4$      \\ \hline
Probability of State: & 1/4 & 1/4 & 1/4 &  1/4   \\ \hline
Value of selecting option $1$:   & $H$ & $H$ & $L$ &  $L$    \\ \hline
Value of selecting option $2$:   & $H$ & $L$ & $H$ & $L$    \\ \hline
\end{tabular}
\end{table}

If both option $1$ and option $2$ have realized their high value $H$, we should expect that the agent is more likely to select option $1$\hypertarget{fo6}{.} Our intuition is that the agent should be more attentive to option $1$, and thus should be more cognisant of option $1$'s high value, and more likely to select it. Similarly, if option $1$ and option $2$ have both realized their low value $L$, then we should expect that the agent is more likely to select option 2.\footnote{Our intuition is that the agent should be more attentive to option $1$, and thus should be more cognisant of option $1$'s low value, and less likely to select it.}

Because of this, if an econometrician, who does not know that the two options have the same value distribution, tried to deduce the two values of option $1$, $H_1$ and $L_1$, and the two values of option $2$, $H_2$ and $L_2$, using a multinomial logit regression, they would decide that $H_1$ is more than the true value $H$, and that $L_1$ is less than the true value $L$ (as is shown rigorously in \hyperlink{sec5}{Section 5}). Fitting thus falls prey to an informational bias, undermining the value of any counterfactual or welfare analysis. 

This type of bias has not previously been identified in the literature on RI. Let $\text{Pr}(n|\omega)$ denote the probability that the agent selects option $n$ in state $\omega$. Let $\text{Pr}(n) = \sum_\omega \text{Pr}(n|\omega) \mu(\omega)$ denote the unconditional probability that option $n$ is selected. \citeA{matmck15} show that fitting of multinomial logit results in the value of an option $n$ to be biased by $\log(\text{Pr}(n)\cdot N)$ in all states $\omega$, where $N$ is the number of available options. The bias found by \citeA{matmck15} can be identified by examining the unconditional choice probabilities of the agent because the driving mechanism is that the cost of learning causes the agent to be biased towards options that they have a higher chance of selecting \textit{a priori}. The bias previously found by \citeA{matmck15} is fundamentally different than the bias demonstrated in this example because their bias does not allow for an option to be over valued in some states and under valued in others, which is in contrast with our setting where option 1 is over valued when it is of high value, and is undervalued when it is of low value. 

An econometrician who observes equal unconditional choice probabilities in this environment, as is predicted in this setting by the model developed in this paper, might be tempted conclude, based on the previous literature, that their analysis is not susceptible to informational biases since each option has the same chance of being selected \textit{a priori}, so the bias of option $n$ is $\log(\text{Pr}(n)\cdot N)= \log(\frac{1}{2}\cdot 2) = 0\,\,\forall n$, and thus any counterfactual or welfare analysis that they conduct is valid. This conclusion may not be correct given the results in this paper.

Further, RU models and RI models with Shannon Entropy can both be rejected for RI with MSSE in this environment if we are able to alter the correlation between the values of the two options. If a RU model describes the agent, then changing the correlation between the values of the two options would not change the choice behavior of the agent. If the behavior of the agent is instead described by MSSE, then changing the correlation between the values of the two options would change the choice behavior of the agent in individual states. This effect is because the total information that can be acquired from learning the value of option 1 (the option that is easier to learn about) changes with the correlation of the options' values. Further, if the above MSSE specification is correct, the unconditional choic\hypertarget{fo7}{e} probabilities of the agent would remain constant when correlation is changed due to the symmetry of the environment, as long as the agent is doing some learning.\footnote{The agent is doing some learning if their choice probabilities differ at all in states of the world that are realized with positive probability.} Finally, if the behavior of the agent is instead described by Shannon Entropy, then the choice behavior in the individual states could only change if the unconditional choice probabilities changed, which is not the case with MSSE. With MSSE, since choice probabilities in a state can be impacted by choice probabilities that are conditioned on some larger subset of states, not only payoffs and unconditional choice probabilities, choice probabilities in a state can change even when payoffs and unconditional choice probabilities do not.

\hypertarget{sec3}{}

\section{Multisource Shannon Entropy (MSSE)}

\hypertarget{sec3.1}{ }

In this section we use axioms to develop this paper's measure of uncertainty. The goal of our axioms are to measure the total amount of uncertainty, which is the expected cost to the agent of perfectly observing the state of the world. The measure of total uncertainty that we develop can then be used to study a rationally inattentive agent because the cost of a noisy information strategy can be taken to be the expected reduction in total uncertainty, as is frequently done with Shannon Entropy in models of RI. Thus, while this paper is interested in studying an inattentive agent that only partially learns about the state of the world, this section discusses an attentive agent that perfectly observes the state of the world.

\subsection{Formal Setting}

As was mentioned in \hyperlink{sec2}{Section 2}, we are interested in an agent who is researching a measurable space $(\Omega, \, \mathcal{F})$. $\Omega$ is a finite set of possible {states of the world}. $\mathcal{F}$ is the set of {events} generated by $\Omega$.

One natural way to think about an agent learning is through a series of questions that have answers that are uniquely determined by the state of the world. These are questions that you can answer if you know the state of the world. How do we model such questions? A \textbf{partition} $\mathcal{P}$ of a state space $\Omega$ is a set of more than one disjoint events in $\mathcal{F}$ whose union is $\Omega$. Notice that our definition of a partition excludes trivial partitions that only contain a single event.

A question with multiple potential answers is thus equivalent to a partition whenever the answer to the question is deterministically determined by the state of the world. This equivalence occurs since every state space we consider has finite possible states of the world, so every such question must have a finite number of answers, and we can simply group states of the world based on the answer to the question they produce. Because we are concerned with questions that have answers that are deterministically determined by the state of the world, the words `question' and `partition' can be used interchangeably.

The simplest kind of question in this setting is a yes or no question. A yes or no question is equivalent to a \textbf{binary partition} $\mathcal{P}^b$ of $\Omega$, which we define as a set of two events, $\mathcal{P}^b = \{A_1,\,A_2\}$, such that $A_1 \cup A_2 = \Omega$, and $A_1 \cap A_2 = \emptyset$. The two phrases `binary partition' and `yes or no question' can thus be used interchangeably.

If $\omega\in\Omega$ is the state of the world, let the \textbf{realized event} of the partition $\mathcal{P}= \{A_1,\,\dots,\,A_m\}$ be denoted by $\mathcal{P}(\omega)$, that is $\mathcal{P}(\omega)=A_i\in \{A_1,\,\dots,\,A_m\}$ iff $\omega\in A_i$. Given a probability measure $\mu:\mathcal{F}\rightarrow \mathbb{R}_+$, and some partition $\mathcal{P}$, let $C(\mathcal{P},\, \mu)$ denote the cost of learning the realized event $\mathcal{P}(\omega)$ of $\mathcal{P}$. $C(\mathcal{P},\, \mu)$, the cost of answering `What is the realized event of $\mathcal{P}$?', given the agent's prior belief, is the basic building block of this paper.

A \textbf{learning strategy}, $S=(\mathcal{P}_1,\,\dots,\,\mathcal{P}_n)$, is a list of partitions whose realized events are successively observed by the agent such that if $\mathcal{P}_i,\,\mathcal{P}_j\in S$, and $i\neq j$, then $\mathcal{P}_i\neq \mathcal{P}_j$. A `learning strategy' is thus `a series of questions', and the two phrases can be used interchangeably. If a learning strategy consists of only binary partitions, we call it a \textbf{binary learning strategy}, and denote it $S^b=(\mathcal{P}_1^b,\,\dots,\,\mathcal{P}_n^b)$. The order of the questions in a learning strategy is important, and changing the order results in a different learning strategy. If, for instance, some questions are more costly for the agent to answer, and help to identify states that are seldom observed, then it may seem efficient for a learning strategy to leave these questions towards the end. The order of the events in a partition, in contrast, is not important, and switching the order in which the events in a partition are listed does not result in a different partition. 

Define $C(S,\,\mu)$, the expected cost of a learning strategy $S=(\mathcal{P}_1,\,\dots,\,\mathcal{P}_n)$, given a probability measure $\mu$, to be the sum of the expected costs of each of the questions in $S$:
$$C(S,\,\mu) = C(\mathcal{P}_1,\,\mu) + \mathlarger{\mathbb{E}}\bigg[ C\Big(\mathcal{P}_2,\,\mu(\cdot|\mathcal{P}_1(\omega))\Big)+\dots+ C\Big(\mathcal{P}_n,\,\mu(\cdot|\mathlarger{\cap}_{i=1}^{n-1}\mathcal{P}_i(\omega))\Big)\bigg].
$$
\noindent Our definition of $C(S,\,\mu)$ thus imposes a form of constant marginal cost onto learning strategies because over the course of their learning strategy the agent does not fatigue, nor do they gain experience with research and become better at learning: all that matters for determining the cost of each question are the beliefs of the agent immediately before the question is answered, and not how much has previously been learned.

If $\mathcal{P}=\{A_1,\,\dots,\,A_m\}$ is a partition, let $\sigma(\mathcal{P})$ denote the $\mathbf{\sigma}$\textbf{-algebra generated by} $\mathcal{P}$, which is the smallest $\sigma$-algebra that contains all the events $A_1,\,\dots,\,A_m$ in $\mathcal{P}$ (which is also the power set of the events in $\mathcal{P}$, since $\mathcal{P}$ is a partition). In general, if $B$ is any collection of partitions, let $\sigma(B)$ denote the $\mathbf{\sigma}$\textbf{-algebra generated by }$ B$, which is the smallest $\sigma$-algebra containing all the events in each of the partitions in $B$. Since a learning strategy $S=(\mathcal{P}_1,\,\dots,\,\mathcal{P}_n)$ is a collection of partitions, we thus use $\sigma(S)$ to denote the ${\sigma}$-algebra generated by $S$.

Sometimes a single question can be as informative as several questions. We say a learning strategy $S$ is \textbf{equivalent} to a partition $\mathcal{P}$ if $\sigma(S) = \sigma(\mathcal{P})$, and we say that a series of questions is equivalent to a particular question if the learning strategy that represents the series of questions is equivalent to the partition that represents the particular question. What $\sigma(S) = \sigma(\mathcal{P})$ means intuitively is that, for any prior probability measure $\mu:\mathcal{F}\rightarrow \mathbb{R}_+$, observing the answers to the series of questions in $S$ always leads to the same posterior as observing the answer to the question `what is the realized event of the partition $\mathcal{P}$?'. We can thus read $\sigma(S) = \sigma(\mathcal{P})$ as saying that, for all priors, $S$ and $\mathcal{P}$ provide the same amount of information to the agent. Let $S(\mathcal{P})\{S|\sigma(S)=\sigma(\mathcal{P})\}$ denote the set of learning strategies that are equivalent to $\mathcal{P}$, and let $S^b(\mathcal{P})=\{S^b|\sigma(S^b)=\sigma(\mathcal{P})\}$ denote the set of binary learning strategies that are equivalent to $\mathcal{P}$.

\hypertarget{sec3.2}{ }
\subsection{Axioms}

What form should a cost function for information take? This difficult question does not have an obvious answer, so this paper takes an axiomatic approach. The axioms make explicit the structure that is imposed on our cost function. Each axiom is meant to be normatively appealing, and can be separately evaluated in different contexts, either empirically, or through introspection, to determine how appropriate it is. Further, the axioms help demonstrate to those that are familiar with Shannon's original axioms \citeyear{sha48} the differences between MSSE and standard Shannon Entropy.

When an agent learns in an inattentive fashion, and only acquires some of the available information, they reduce the amount that remains to be learned, and thus reduce the subsequent cost of learning the state of the world. The cost of the inattentive learning done by the agent can thus simple be measured as the reduction in the cost of learning the state of the world, as subsequent sections discuss,\footnote{We argue later in the paper that the application of Shannon Entropy can be interpreted in this same fashion.} as long as we can establish the cost of learning the state of the world for different probability measures.

Thus, while the learning of an agent is frequently inattentive, and this paper wishes to study environments where the agent only partially learns about the state of the world, this section discusses an attentive agent that tries to perfectly observes the state of the world. We do this because we want our axioms to be normatively appealing, and we find axioms about perfectly observing the state of the world to be a more intuitive, and hence easier to evaluate normatively, than axioms that focus directly on inattentive behavior, describing costs of different kinds of stochastic experiments. Another interpretation of this strategy is that, while the primitive of our model is the cost of learning the realized events of partitions, and the agent could choose to learn through such partitions of the state space, we do not constrain the agent's choice of information strategy so that they must learn through such partitions, and they can instead choose a noisy signal structure if they desire.

We now state the five axioms required to achieve this paper's measure of uncertainty, the cost of perfectly learning the state of the world:

\bigskip

\hypertarget{ax1}{}
\noindent \textbf{Axiom 1 (Efficient Yes or No Questions):} Given a partition $\mathcal{P}$, for all probability measures $\mu$:
$$C(\mathcal{P},\, \mu) \geq \min\limits_{S^b\in S^b(\mathcal{P})} C(S^b,\,\mu).
$$

\bigskip

In plain language, \hyperlink{ax1}{Axiom 1} says that for any question $\mathcal{P}$, there are a series of yes or no question $S^b$, that provide the same amount of information as $\mathcal{P}$, and can be asked instead for the same cost or less. This assertion allows us to focus on series of yes or no questions without loss when we try to determine the cost to the agent of learning the state of the world, which is supported by research in the psychology and psychophysics literatures. 

Eye tracking analysis shows that when agents are faced with multiple options, they successively compare pairs of the options along a single attribute dimension \cite{nogste14, nogste18}. This suggests that, in practice, agents are breaking their learning into a number of smaller queries. Further, in the psychology literature these pairwise comparisons are frequently modelled as ordinal in nature \cite{nogste18}, equivalent to questions with binary outcomes, e.g. `Is option $a$ better than option $b$ in dimension $x$?', instead of more complicated questions, e.g. `How much better is option $a$ than option $b$ in dimension $x$?', because findings in the field of psychophysics suggest that agents are good at discriminating stimuli, but are not good at determining the magnitude of the same stimuli \cite{steea06}.

Before we introduce the rest of our axioms, we pause to discuss learning strategy invariance, a concept that helps us to make it explicit what we are assuming with the rest of our axioms. In general, a particular question $\mathcal{P}$, and an equivalent series of questions $S$, may produce different expected costs depending on what questions are selected, and how they are ordered in $S$. A given question $\mathcal{P}$, however, may have the peculiar property that, given any prior, all series of questions that are equivalent to it have the same expected cost. If a question has this strong property, we say it is learning strategy invariant. Formally, we say a partition $\mathcal{P}$ is \textbf{learning strategy invariant}, if for each probability measure $\mu$, the expected cost $C(S,\,\mu)$ is the same for every learning strategy $S$ that is equivalent to $\mathcal{P}$. 

In many environments there are questions that are not learning strategy invariant. Consider the environment described in \hyperlink{sec2.2}{Example 2} in \hyperlink{sec2.2}{Section 2.2}. In this context, let $A_1 = \{\omega_1,\,\omega_2\}$, $A_2 = \{\omega_1,\,\omega_3\}$, $\mathcal{P}_1^b=\{A_1,\,A_1^c\}$, and $\mathcal{P}_2^b=\{A_2,\,A_2^c\}$. Notice that observing the realized event of $\mathcal{P}_1^b$ is equivalent to learning the value of option 1, and observing the realized event of $\mathcal{P}_2^b$ is equivalent to learning the value of option 2. Now, let $\mathcal{P}_3 = \{\{\omega_1\},\,\{\omega_2\},\,\{\omega_3\},\,\{\omega_4\}\}$ denote our partition of the state space. Notice that the learning strategy $S^b = (\mathcal{P}_1^b,\,\mathcal{P}_2^b)$ is equivalent to $\mathcal{P}_3$, because if we answer `What is the value of option 1?', and then answer `What is the value of option 2?', we have observed the state of the world. 

Based on our discussion in \hyperlink{sec2.2}{Section 2.2}, however, we should expect that $\mathcal{P}_3$ may not be learning strategy invariant. Consider $\tilde{S}^b = (\mathcal{P}_2^b,\,\mathcal{P}_1^b)$, which is also equivalent to $\mathcal{P}_3$. If the value of option 1 and option 2 were perfectly correlated, then observing the value of one of them would tell you the value of the other. The cost of $S^b$ would then be the cost of observing the value of option 1, which we assumed to be less than the cost of observing the value of option 2, which is then the cost of $\tilde{S}^b$. 

A set of partitions that are certainly learning strategy invariant, in contrast, is the set of binary partitions. If $\mathcal{P}^b$ is a binary partition, then $\mathcal{P}^b$ is learning strategy invariant because the only learning strategy $S$ such that $\sigma(S) = \sigma(\mathcal{P}^b)$, is $S=(\mathcal{P}^b)$. Thus, for any $\mu$, all learning strategies $S$ such that $\sigma(S) = \sigma(\mathcal{P}^b)$ have the same expected cost $C(S,\,\mu)=C(\mathcal{P}^b,\,\mu)$.

\hyperlink{ax1}{Axiom 1} means that it is without loss for us to think about an agent learning through binary partitions. Since binary partitions are learning strategy invariant, it is thus without loss for us to consider the agent learning through learning strategy invariant partitions. As such, our four remaining axioms are only concerned with the costs of questions that are learning strategy invariant. These axioms are thus rather weak in nature, only imposing structure onto the costs of a particular kind of question that already features a lot of structure. We eventfully interpret different earning strategy invariant partitions as corresponding to different information sources, hence the name MSSE.

\bigskip

\hypertarget{ax2}{}
\noindent \textbf{Axiom 2 (More States are Harder to Differentiate):} If a partition $\mathcal{P}$ is a learning strategy invariant, and $\mu$ is a probability measure such that $n$ events $\{A_i\}_{i=1}^n \subset \mathcal{P}$ are given probability 1/$n$, while $\tilde{\mu}$ is a probability measure such that $n+1$ events $\{B_j\}_{j=1}^{n+1}\subseteq \mathcal{P}$ are given probability 1/$(n+1)$, then $C(\mathcal{P},\,\mu) < C(\mathcal{P},\,\tilde{\mu}).$

\bigskip

In plain language, \hyperlink{ax2}{Axiom 2} says that it is cheaper to use a given learning strategy invariant partition to differentiate between fewer more likely states than it is to differentiate between more less likely states.

\bigskip

\hypertarget{ax3}{}
\noindent \textbf{Axiom 3 (Continuity):} If a partition $\mathcal{P}$ is learning strategy invariant, there are two probability measures $\mu$ and $\tilde{\mu}$ assign positive probability to the same events in $\mathcal{P}$, and $\mu_\alpha$ is defined for each $\alpha\in[0,\,1]$ so that $\mu_\alpha(\omega)= \alpha \mu (\omega) + (1-\alpha)\tilde{\mu}(\omega)$ for all $\omega\in \Omega$, then $C(\mathcal{P},\,\mu_\alpha)$ changes continuously in $\alpha$.

\bigskip

In plain language, \hyperlink{ax3}{Axiom 3} says that if the events that are given a positive probability of occurring in a learning strategy invariant partition $\mathcal{P}$ do not change, then the cost of learning which event in $\mathcal{P}$ has occurred, $C(\mathcal{P},\,\mu)$, should change continuously with respect to $\mu$. This is intuitive since small changes in the chances of events occurring should not result in a large change in the cost of a question that differentiates between said events, if which events are possible does not change.

\bigskip

\hypertarget{ax4}{}
\noindent \textbf{Axiom 4 (Measurement):} If a partition $\mathcal{P}$ is learning strategy invariant, then $C(\mathcal{P},\,\mu)$ is measurable with respect to $\sigma(\mathcal{P})$.

\bigskip

In plain language, \hyperlink{ax4}{Axiom 4} says that if $\mathcal{P}$ is learning strategy invariant, then the expected cost of the question represented by $\mathcal{P}$ should be fully determined by the chance that each of its answers occurs. If we know the question that is being asked, and the the chance of each of its answers, then we know the expected cost of asking the question, we do not require any additional information.

Our next axiom, \hyperlink{ax5}{Axiom 5}, addresses the subjective nature of the state space. To ease exposition, we first introduce some new definitions. A finite state space $\Omega$ is \textbf{coarser} than another finite state space $\tilde{\Omega}$, if each state in $\Omega$ corresponds to an event in the set of events generated by $\tilde{\Omega}$. In other words, a finite state space $\Omega$ is {coarser} than another finite state space $\tilde{\Omega}$, if each state in $\Omega$ corresponds to a union of states in $\tilde{\Omega}$. We say the state space $\tilde{\Omega}$ is \textbf{finer} than $\Omega$, if $\Omega$ is {coarser} than $\tilde{\Omega}$. Similarly, a partition $\mathcal{P}$ of a state space $\Omega$ is \textbf{coarser} than a partition $\tilde{\mathcal{P}}$ of the same state space $\Omega$, if each event in $\mathcal{P}$ corresponds to a union of events in $\tilde{\mathcal{P}}$. We say a partition $\tilde{\mathcal{P}}$ is \textbf{finer} than a partition $\mathcal{P}$ if $\mathcal{P}$ is \textbf{coarser} than $\tilde{\mathcal{P}}$. Notice that if $\mathcal{P}$ is a partition of a state space $\Omega$, and a state space $\tilde{\Omega}$ is finer than $\Omega$, then $\mathcal{P}$ is also a partition of $\tilde{\Omega}$, since each event in $\mathcal{P}$ is an event in the event space generated by both $\Omega$ and $\tilde{\Omega}$.

In practice, `the' state space $\Omega$ is determined by the researcher and the application, and referring to it as `the' state space is typically a misnomer. Sometimes the researcher uses just enough states so that the payoff function is measurable, as in \hyperlink{sec2.2}{Example 2}, and other times the researcher includes more states than are required for measuring the payoff function because it is deemed relevant to the agent, as in \hyperlink{sec2.1}{Example 1}. In \hyperlink{sec2.1}{Example 1}, however, the four `states' we describe are not actually states, they are events in some richer underlying state space. Two realizations of the `state' of the world in which $51$ blue balls appear may differ from each other because the balls may appear in a different order, which may be relevant for the agent's cost of learning.\footnote{Imagine that the 100 balls from \hyperlink{sec2.1}{Example 1} are displayed on the screen in ten rows of ten, and that 51 are blue. If the position of the red and blue balls appears random, it would seem intuitive that it is more costly for the agent to learn whether or not the the majority of the ball are blue compared to the setting where the top five rows all consist of ten blue balls, and in the bottom ten rows there is one blue ball and the rest are red.} Further, a change in the payoff function may necessitate the description of a richer state space. While our description of the state space can change, the reality of the agent does not, and the cost of asking certain questions should not change based on the subjective modelling decisions of the researcher. Thus, when we consider the cost of learning the outcomes of a learning strategy invariant partition $\mathcal{P}$, our cost should not reject the potential that there is another learning strategy invariant partition $\tilde{\mathcal{P}}$ of a more detailed finite state space $\tilde{\Omega}$. This notion is summarized in \hyperlink{ax5}{Axiom 5}.

\bigskip

\hypertarget{ax5}{}
\noindent \textbf{Axiom 5 (Subdivision):} If a partition $\mathcal{P}=\{A_1,\,\dots,\,A_m\}$ of a state space $\Omega$ is learning strategy invariant, then for any $n>m$, $C(\mathcal{P},\,\mu)$ does not rule out that there is a finer state space $\tilde{\Omega}$, of which there is a learning strategy invariant partition $\tilde{\mathcal{P}}=\{\tilde{A}_1,\,\dots,\,\tilde{A}_n\}$, which is finer than $\mathcal{P}$.

\bigskip

In plain language, \hyperlink{ax5}{Axiom 5} says that we do not want to impose a cost function onto a limited dataset that rules out finer learning strategy invariant partitions on a more detailed finite state space. \hyperlink{ax5}{Axiom 5} is important because it allows us to consider learning strategy invariant partitions with arbitrarily many events, even when our state space $\Omega$ has relatively few states, which is crucial for the proof of \hyperlink{th1}{Theorem 1}.

The first result that we show using our axioms is that if $\mathcal{P}$ is a leaning strategy invariant partition, then $C(\mathcal{P},\,\mu)$ is constant with respect to permutations of the probability measure $\mu$. If $\mathcal{P}=\{A_1,\,\dots,\,  A_m\}$ is a leaning strategy invariant partition, we say that $\tilde{\mu}$ is a \textbf{permutation} of $\mu$ on $\mathcal{P}$ if there is a bijection $\pi:\{1,\,\dots,\,m\}\rightarrow \{1,\,\dots,\,m\}$ such that $\forall i\in \{1,\,\dots,\,m\},\,\mu(A_i) = \tilde{\mu}(A_{\pi(i)}).$

\bigskip

\hypertarget{le1}{}
\noindent \textbf{Lemma 1.} If a partition $\mathcal{P}$ is learning strategy invariant, and $C$ satisfies \hyperlink{ax4}{Axiom 4}, and \hyperlink{ax5}{Axiom 5}, then if $\tilde{\mu}$ is a {permutation} of $\mu$ on $\mathcal{P}$, $C(\mathcal{P},\,\mu) = C(\mathcal{P},\,\tilde{\mu}).
$

\smallskip

\noindent Proofs for results in \hyperlink{sec3}{Section 3} and can be found in \hyperlink{ap1}{Appendix 1}

\bigskip

The second result that we show using our axioms helps demonstrate the difference between MSSE and standard Shannon Entropy.

\bigskip

\hypertarget{le2}{}
\noindent \textbf{Lemma 2.} If a partition $\mathcal{P}$ is learning strategy invariant, and $C$ satisfies \hyperlink{ax2}{Axiom 2}, \hyperlink{ax3}{Axiom 3}, \hyperlink{ax4}{Axiom 4}, and \hyperlink{ax5}{Axiom 5,} then there exists a multiplier $\lambda(\mathcal{P})\in\mathbb{R}_{++}$, such that for all probability measures $\mu$: $C(\mathcal{P},\,\mu)= \lambda(\mathcal{P}) \mathcal{H}(\mathcal{P},\,\mu),$
\noindent where $\mathcal{H}$ is Shannon's standard measure of entropy \citeyear{sha48}, defined in equation (\hyperlink{1}{1}).

\bigskip

Underlying each learning strategy invariant partition is some information source that allows the agent to differentiate between the events that comprise the partition. \citeA{sha48} imposes learning strategy invariance onto all partitions of $\Omega$, which implies that all partitions have the same costs associated with them (there is a $\lambda>0$ such that $\lambda(\mathcal{P})=\lambda$ for all partitions $\mathcal{P}$ of $\Omega$), and so it is without loss to think of the agent as learning from a single information source that allows them to differentiate between the different states of the world. With MSSE, in contrast, different learning strategy invariant partitions are allowed to have different costs associated with them ($\lambda(\mathcal{P})$ may differ depending on the learning strategy invariant partition $\mathcal{P}$), and thus it is natural to think of the agent as learning different pieces of information from different sources depending on which source allows them to acquire the information at the lowest costs, as is formalized by \hyperlink{th1}{Theorem 1}. This interpretation is how MSSE gets its name.

\hypertarget{sec3.3}{ }
\subsection{Total Uncertainty}
 
\hyperlink{le2}{Lemma 2} tells us that for each binary partition $\mathcal{P}^b$, there is an \textbf{associated multiplier}, $\lambda(\mathcal{P}^b)\in\mathbb{R}_{++}$, such that for all probability measures $\mu$: $C(\mathcal{P}^b,\,\mu) = \lambda(\mathcal{P}^b)\mathcal{H}(\mathcal{P}^b,\,\mu)$. Since there are a finite number of binary partitions of $\Omega$, we can order the binary partitions by their associated multipliers. Let $\lambda_1$ denote the multiplier associated with all binary partitions, denoted $\{\mathcal{P}_i^{b,\lambda_1}\}_{i=1}^{n_1}$, with the lowest multiplier. 

If the agent can always learn the state of the world by asking questions with multiplier $\lambda_1$, then $\sigma(\{\mathcal{P}_i^{b,\lambda_1}\}_{i=1}^{n_1})=\mathcal{F}$, and we let $M$=1.\footnote{If $M$=1, then MSSE collapses to standard Shannon Entropy.} If not, let $\lambda_2$ denote the multiplier associated with all binary partitions, denoted $\{\mathcal{P}_i^{b,\lambda_2}\}_{i=1}^{n_2}$, with the second lowest multiplier.

If the agent can always learn the state of the world by asking questions with multipliers $\lambda_1$ or $\lambda_2$, then $\sigma(\{\mathcal{P}_i^{b,\lambda_1}\}_{i=1}^{n_1},\,\{\mathcal{P}_i^{b,\lambda_2}\}_{i=1}^{n_2})=\mathcal{F}$, and we let $M=2$. If not, let $\lambda_3$ denote the multiplier associated with all binary partitions, denoted $\{\mathcal{P}_i^{b,\lambda_3}\}_{i=1}^{n_3}$, with the third lowest multiplier.

Continue in this fashion until we let $\lambda_M$ denote the multiplier associated with all binary partitions, denoted $\{\mathcal{P}_i^{b,\lambda_M}\}_{i=1}^{n_M}$, with the lowest multiplier such that the state of the world is always revealed when all questions with equal or lower associated multipliers are asked, that is, the lowest $M$ such that: $\sigma(\{\mathcal{P}_i^{b,\lambda_1}\}_{i=1}^{n_1},\,\dots,\,\{\mathcal{P}_i^{b,\lambda_M}\}_{i=1}^{n_M})=\mathcal{F}$.

To help make our notation more compact, we can use a group of partitions to \textbf{generate} a finer partition: if $(\mathcal{P}_1,\,\dots,\,\mathcal{P}_m)$ is a group of partitions, let $\times\{\mathcal{P}_i\}_{i=1}^n$ denote the partition such that $\sigma(\times\{\mathcal{P}_i\}_{i=1}^n)=\sigma(\mathcal{P}_1,\,\dots,\,\mathcal{P}_n)$. Then, for $j\in\{1,\,\dots,\,M\}$,\footnote{$M$ is defined in the proceeding paragraphs.} let $\mathcal{P}_{\lambda_j}=\times\{\mathcal{P}_i^{b,\lambda_j}\}_{i=1}^{n_j}$.

MSSE incorporates different perceptual distances because it allows for different events to be different distances from each other. Events in $\mathcal{P}_{\lambda_1}$, for instance, have greater perceptual distances between them than events in $\mathcal{P}_{\lambda_M}$ (assuming $M>1$).

Since $\Omega$ is a partition of itself, we can, as a minor abuse of notation, let $S(\Omega)=\{S|\sigma(S)=\mathcal{F}\}$ denote the set of learning strategies such that $\sigma(S)=\sigma(\Omega)=\mathcal{F}$.

\bigskip
\hypertarget{th1}{}
\noindent \textbf{Theorem 1.} If $C$ satisfies all five axioms, then there exist partitions $\mathcal{P}_{\lambda_1},\,\dots,\, \mathcal{P}_{\lambda_M}$ as defined above, and constants $\lambda_1<\,\dots\,<\lambda_M$, such that for any probability measure $\mu$ on $\mathcal{F}$
$$ \min\limits_{S\in S(\Omega)} C(S,\,\mu) =\lambda_\mathsmaller{1} \mathcal{H}\Big(\mathcal{P}_{\lambda_1},\,\mu\Big)+ \mathlarger{\mathbb{E}}\bigg[ \lambda_\mathsmaller{2}\mathcal{H}\Big(\mathcal{P}_{\lambda_2},\,\mu(\cdot|\mathcal{P}_{\lambda_1}(\omega))\Big)+\dots+\lambda_\mathsmaller{M} \mathcal{H}\Big(\mathcal{P}_{\lambda_M},\, \mu(\cdot|\mathlarger{\cap}_{{i=1}}^{{M-1}}\mathcal{P}_{\lambda_i}(\omega))\Big)\bigg],
$$
\noindent where $\mathcal{H}$ is defined as in equation (\hyperlink{1}{1}).

\bigskip

In plain language, \hyperlink{th1}{Theorem 1} says that if the cost of learning satisfies all five axioms, then the cheapest way (in expectation) to learn the state of the world always involves first asking all the yes or no questions with the lowest associated multiplier (in any order)\hypertarget{fo5}{,} then asking all the yes or no questions with the second lowest multiplier, and continuing in this fashion until the state of the world has been realized.

\hyperlink{th1}{Theorem 1} generates the more flexible measure of uncertainty that we desired for studying inattentive behavior. If the agent starts with a prior $\mu$, and does optimal learning that reaches a posterior $\tilde{\mu}$, then we let the cost of this inattentive research be the reduction in the cost of perfectly learning the state of the world, as is discussed in the next section.

In terms of Shannon's original context, this paper's model can be thought of as describing learning of information from $M$ sources, where source $i$, for $i\in\{1,\,2,\dots,\,M\}$, is capable of providing information about $\mathcal{P}_{\lambda_i}(\omega)$. Shannon's original axioms, in contrast, impose that all partitions $\mathcal{P}$ are learning strategy invariant, which is analogous to all binary partitions having the lowest multiplier, and there only being one information source relevant for learning.

The $\mathcal{P}_{\lambda_i}$'s that could be used in \hyperlink{th1}{Theorem 1} are not unique, with the exception of $\mathcal{P}_{\lambda_1}$. The versions described in the paragraphs preceding \hyperlink{th1}{Theorem 1} are the unique coarsest partitions that could be used in the statement of the theorem. For $i\in\{2,\,\dots,\,M\}$, $\mathcal{P}_{\lambda_i}$ could, for instance, be replaced by $\tilde{\mathcal{P}}_{\lambda_i}=\times\{\mathcal{P}_{\lambda_j}\}_{j=1}^{i}$ in the statement of \hyperlink{th1}{Theorem 1}, which would constitute the unique finest representation of the partitions.

The axiomatic derivation of the cost benchmark in this paper requires a discrete state space for the state of the world, as is the case with Shannon Entropy. If a continuous state space is desired for the state of the world, however, a measure of uncertainty for a continuous state space can be defined in an analogous manner to the measure of uncertainty defined in \hyperlink{th1}{Theorem 1} for a discrete state space, which is similar to what is done by \citeA{sha48} to apply Shannon Entropy in a continuous setting.

\hypertarget{sec4}{}
\section{Inattentive Learning with MSSE}

The following section introduces and solves a model of RI that uses MSSE to measure the cost of acquiring information. We establish that our new more flexible measure of uncertainty can still be incorporated tractably into a model of RI, which is not an obvious result. Apart from the use of MSSE instead of Shannon Entropy for the measurement of uncertainty, this section follows the work of \citeA{matmck15} closely so as to aid comparison between the two models.

Given our result in \hyperlink{th1}{Theorem 1}, we take the expected cost of a particular information strategy to be defined as:
$$\textbf{C}(F(s,\,\omega),\,\mu)   =\mathbb{E}\bigg[\min\limits_{S\in S(\Omega)} C(S,\,\mu) -\min\limits_{S\in S(\Omega)} C(S,\,\mu(\cdot|s))\bigg].$$
\noindent A noisy information strategy reduces the total amount of uncertainty, and we thus measure the cost of such a noisy information strategy as the expected reduction in total uncertainty. This interpretation can also be applied to RI models that use Shannon Entropy to measure the cost of noisy information structures. Shannon Entropy is a measure of total uncertainty derived from axioms about the cost of successively learning the realized events of partitions, and in such models the cost of a noisy signal is simply taken to be the reduction in total uncertainty, as measured by Shannon Entropy.

The cost functions that can be defined as above with MSSE are in the class of uniformly posterior-separable cost functions described by \citeA{capea17}. The behavior generated in static settings by such posterior-separable cost functions has been shown to be equivalent to the behavior generated by sequential information sampling in some dynamic contexts \cite{hebwoo17, morstr19}. In particular, \citeA{hebwoo17} show that a class of static cost functions, which they call `neighborhood-based' cost functions, can be micro-founded in this way. The cost functions explored in this paper that measure the reduction in MSSE are a strict subset of the neighborhood-based cost functions described in their paper, and thus the cost functions in this paper are micro-founded in two ways, directly through the axioms in this paper, and indirectly through the dynamic analysis conducted by \citeA{hebwoo17}. While symmetry imposes a unique set of partitions in \hyperlink{sec2.1}{Example 1} when MSSE is used, there are numerous representations that can be used when a neighborhood-based cost function is assumed. \citeA{hebwoo17} suggest two ways of modelling the neighborhoods in such a setting, one of which is fitted by \citeA{deanel17}, and neither of which is equivalent to the partitions suggested by MSSE. 

\citeA{hueea19}, in turn, create an ad hoc group of cost functions that are also a generalization of Shannon Entropy, but are a strict subset of the cost functions studied in this paper that measure reduction in MSSE. The cost functions studied by \citeA{hueea19} allow for multiple perceptual distances, but are not capable of predicting the behavior we argued was intuitive in \hyperlink{sec2.1}{Example 1}, since in \hyperlink{sec2.1}{Example 1} their cost functions collapses to standard Shannon Entropy.

\hypertarget{sec4.1}{ }

\subsection{Rationally Inattentive Agent's Problem}

As was discussed in \hyperlink{sec2}{Section 2}, when the agent faces a probability space $(\Omega,\,\mathcal{F},\, \mu)$ and a set of options $N$, the agent's problem is to maximize the expected value of the option they select less the cost of learning by choosing an optimal {information strategy}, and subsequently selecting an option based on the signal produced by their information strategy. The agent's problem can thus be written:
\hypertarget{2}{}
\begin{equation}
\max_{F\in \Delta(\mathbb{R} \times \Omega)} \sum\limits_{\omega\in\Omega}\int\limits_s V(s|F)F(ds|\omega)\mu(\omega) - \textbf{C}(F(s,\,\omega),\,\mu),
\end{equation} 
\hypertarget{3}{}
\begin{equation}
\text{such that } \forall \omega \in \Omega:\, \int\limits_s F(ds,\,\omega) = \mu(\omega).
\end{equation}

The above problem is complicated and not particularly tractable, so we follow \citeA{matmck15} and re-write this problem directly in terms of the choice probabilities of the agent. This process requires the development of some new notation. Define $S(n|F)=\{s\in\mathbb{R}:\,F(s)>0,\,a(s|F)=n\}$, to be the set of signals that result in the agent selecting option $n$. Next, as was done in \hyperlink{sec2}{Section 2}, define the chance of option $n$ being selected conditional on the state of the world to be:
\hypertarget{4}{}
\begin{equation}
\text{Pr}(n|\omega) = \int\limits_{s\in S(n|F)} F(ds|\omega),
\end{equation} 
\noindent and for event $A\in\mathcal{F}$, define the chance of $n$ being selected conditional on $A$ being realized to be:
\hypertarget{5}{}
\begin{equation}
\text{Pr}(n|A) = \sum\limits_{\omega \in A} \text{Pr}(n|\omega) \mu(\omega|A).
\end{equation} 
\noindent Define the \textbf{unconditional choice probability} of option $n$ to be: 
\hypertarget{6}{}
\begin{equation}
\text{Pr}(n) = \sum\limits_{\omega \in \Omega} \text{Pr}(n|\omega) \mu(\omega).
\end{equation}
\noindent Denote the collection $\{\text{Pr}(n|\omega)\}_{n=1}^N$ by $\mathbb{P}$. Using this notation, we can re-write the agent's problem:

\bigskip

\hypertarget{le3}{}

\noindent \textbf{Lemma 3.} Choice probabilities $\mathbb{P}$ are the outcome of a solution to the agent's problem in (\hyperlink{2}{2}) subject to (\hyperlink{3}{3}) iff they solve:
\hypertarget{7}{}
\begin{equation}
\max_{\mathbb{P}} \sum\limits_{n\in\mathcal{N}} \sum\limits_{\omega\in\Omega} \mathbf{v}_n(\omega) \text{Pr}(n|\omega) \mu(\omega) - \textbf{C}(\mathbb{P},\,\mu),
\end{equation}
\hypertarget{8}{}
\begin{equation}
\text{such that:  } \forall \,n\in\mathcal{N}, \,\, \text{Pr}(n|\omega) \geq 0, \,\,\forall \, \omega\in \Omega,
\end{equation}
\hypertarget{9}{}
\begin{equation}
\text{and}\,\, \sum\limits_{n\in\mathcal{N}} \text{Pr}(n|\omega) =1 \,\, \forall\, \omega\in \Omega,
\end{equation}
\noindent where $\textbf{C}(\mathbb{P},\,\mu)$ is as defined in \hyperlink{le9}{Lemma 9}.

\smallskip

\noindent Proofs for results in \hyperlink{sec4}{Section 4} and \hyperlink{sec5}{Section 5} can be found in \hyperlink{ap2}{Appendix 2}

\bigskip

\noindent This new problem, where the agent selects their conditional choice behavior $\mathbb{P}$, is substantially easier to solve than the problem where the agent picks their information strategy $F(s,\,\omega)$.

\hypertarget{sec4.2}{ }

\subsection{Behavior of a Rationally Inattentive Agent}

Using \hyperlink{le3}{Lemma 3}, we can establish a necessary condition for the optimal behavior of the agent with \hyperlink{th2}{Theorem 2}, and then use said necessary condition to simplify the maximization problem undertaken by the agent with \hyperlink{co1}{Corollary 1}.

\hypertarget{th2}{}
\noindent \textbf{Theorem 2:}

If $\mathbb{P}$ is the solution to (\hyperlink{7}{7}) subject to (\hyperlink{8}{8}) and (\hyperlink{9}{9}), then $\forall \,n\in\mathcal{N}$, and $\forall \,\omega\in\Omega$, the probability that option $n$ is selected in state $w$ satisfies:

\hypertarget{10}{}
\begin{equation}
\text{Pr}(n|\omega) = \dfrac{\text{Pr}(n)^{\frac{\lambda_1}{\lambda_M}}\text{Pr}(n|\mathcal{P}_{\lambda_1}(\omega) )^\frac{\lambda_2 - \lambda_1}{\lambda_M}\dots\,\text{Pr}(n|\cap_{i=1}^{M-1}\mathcal{P}_{\lambda_i}(\omega))^\frac{\lambda_M - \lambda_{M-1}}{\lambda_M} \mathlarger{\mathlarger{e}} ^\frac{\mathbf{v}_n(\omega)}{\lambda_M}}{\mathlarger{\sum}\limits_{\nu\in\mathcal{N}}    \text{Pr}(\nu)^{\frac{\lambda_1}{\lambda_M}} \text{Pr}(\nu|\mathcal{P}_{\lambda_1}(\omega) )^\frac{\lambda_2 - \lambda_1}{\lambda_M}\dots\,\text{Pr}(\nu|\cap_{i=1}^{M-1}\mathcal{P}_{\lambda_i}(\omega))^\frac{\lambda_M - \lambda_{N-1}}{\lambda_M} \mathlarger{\mathlarger{e}} ^\frac{\mathbf{v}_\nu(\omega)}{\lambda_M}}.
\end{equation} 

\bigskip

Those familiar with the work of \citeA{matmck15} will recognize the above formula as the MSSE analogue of \citeA{matmck15}'s Theorem 1. When all partitions are learning strategy invariant, $\lambda_1=\lambda_2=\dots=\lambda_M$, and the above formula collapses to \citeA{matmck15}'s Theorem 1. 

With standard Shannon Entropy, the chance that the agent selects an option thus depends only on the unconditional chances of the options being selected, and the realized values of the options. With MSSE, in contrast, as the above formula indicates, the chance that the agent selects an option $n$ in a particular state of the world $\omega$ depends on the unconditional chances of the options being selected, $\text{Pr}(n)$, the realized values of the options $\textbf{v}_n(\omega)$, as well as the probabilities of the options being selected in similar states of the world. Here `similar states of the world' refers to states that induce the same realization of partitions with associated multipliers smaller than $\lambda_M$. It makes sense that when easier to observe pieces of information indicate that an option $n$ is likely of above average value, that the agent should select option $n$ with a higher probability, even if the above average value has not been realized. For a more complete discussion of the intuitive properties of the choice behavior described in \hyperlink{th2}{Theorem 2}, please see \hyperlink{ap3}{Appendix 3}.

Behavior that is consistent with \hyperlink{th2}{Theorem 2} is not necessarily optimal because in many settings it is not optimal for the agent to consider all of the available options (choose them with positive probability), and though such a corner solution may be optimal, there are many corners that are consistent with \hyperlink{th2}{Theorem 2} but are not optimal. For instance, for any $n\in \mathcal{N}$, if the agent selects $n$ with probability one in all states of the world, then their behavior is consistent with \hyperlink{th2}{Theorem 2}, but it is easy to come up with examples where this would not be optimal for any $n$.

\bigskip

\hypertarget{co1}{}
\noindent \textbf{Corollary 1:} 

Conditional and unconditional choice probabilities described in (\hyperlink{5}{5}) and (\hyperlink{6}{6}) are a solution to (\hyperlink{7}{7}) subject to (\hyperlink{8}{8}) and (\hyperlink{9}{9}) iff they comply with \hyperlink{th2}{Theorem 2} and solve:
$$\max\limits_{\mathbb{P}} \sum\limits_{\omega\in\Omega} \log \Bigg( \sum\limits_{n\in\mathcal{N}} \text{Pr}(n)^{\frac{\lambda_1}{\lambda_M}}\text{Pr}(n|\mathcal{P}_{\lambda_1}(\omega))^\frac{\lambda_2 - \lambda_1}{\lambda_M}\dots\,\text{Pr}(n|\cap_{i=1}^{M-1}\mathcal{P}_{\lambda_i}(\omega))^\frac{\lambda_M - \lambda_{M-1}}{\lambda_M} \mathlarger{\mathlarger{e}} ^\frac{\mathbf{v}_n(\omega)}{\lambda_M}  \Bigg)\mu(\omega),
$$
\noindent such that:
$$\forall \, A\in \mathcal{F}:\,\, \text{Pr}(n|A) \geq 0 \,\,\forall\,n, \,\,\,\,\, \text{and} \,\,\,\,\, \sum\limits_{n\in\mathcal{N}} \text{Pr}(n|A)=1.
$$

\bigskip

\hyperlink{co1}{Corollary 1} is helpful because it reduces the number of choice variables faced by the agent, which means it is easier for the researcher to find optimal agent behavior. When solving the problem described in \hyperlink{le3}{Lemma 3}, the agent must choose $\text{Pr}(n|\omega)$ for all $n$ and $\omega$. When solving the problem in \hyperlink{co1}{Corollary 1}, the agent must only choose $\text{Pr}(n|A)$ for all $n$ and $A\in\times\{\mathcal{P}_{\lambda_i}\}_{i=1}^{M-1}$, which is a coarser partition. In \hyperlink{sec2.2}{Example 2}, for instance, if the agent tries to solve \hyperlink{le3}{Lemma 3} they must pick their probabilities of selecting option 1 and option 2 in four different states of the world, while if they solve the problem in \hyperlink{co1}{Corollary 1} they must only pick their probabilities of selecting option 1 and option 2 in two events, and then \hyperlink{th2}{Theorem 2} dictates their choice probabilities in each state of the world. This reduction makes finding optimal behavior of the agent easier for the researcher because there are thus half as many choice variables when analysing \hyperlink{sec2.2}{Example 2} if \hyperlink{co1}{Corollary 1} is used instead of \hyperlink{le3}{Lemma 3}.

Any choice behavior that complies with \hyperlink{co1}{Corollary 1} and \hyperlink{th2}{Theorem 2} is optimal. This paper does not provide conditions for optimal behavior that are both necessary and sufficient, however, as is done by \citeA{capea16} in a setting with standard Shannon Entropy. This may cause some to view finding optimal behavior with MSSE quite daunting. The necessary and sufficient conditions given by \citeA{capea16} in the setting with Shannon Entropy are appealing because they verify if behavior that satisfies the necessary conditions are in fact optimal, and provide insight into the formation of the agent's optimal consideration set, which is interesting in and of itself. In the more complicated setting studied in this paper, the necessary and sufficient conditions are less appealing. The reality is that almost any problem in this more complicated setting requires a computer for finding optimal behavior, but that is true in the standard setting as well, even with the conditions derived by \citeA{capea16}. The good news is that the optimization problem that needs to be solved involves maximization of a strictly concave function over a compact domain, which is differentiable everywhere on the interior. Thus, standard steepest accent algorithms work well for solving the problem described in \hyperlink{co1}{Corollary 1}, even when the number of options in $\mathcal{N}$ and the number of events in $\times\{\mathcal{P}_{\lambda_i}\}_{i=1}^{M-1}$ are large. For those that are interested in a discussion of how MSSE changes the formation of optimal consideration sets, please see \citeA{wal19b}. Further, while \citeA{hueea19} do attempt to provide necessary and sufficient conditions for optimal behavior in their setting, the conditions are incorrect, as is also discussed by \citeA{wal19b}.

As is true in the setting with standard Shannon Entropy, optimal choice behavior may not be unique. If two options are known \textit{a priori} to take the same value in each state of the world, for instance, then the agent can shift probability from one of these two options to the other whenever the former has a strictly positive probability of being selected in an optimal solution. While these sorts of environments are possible, generically optimal behavior is unique. This feature of optimal behavior should be evident since payoffs are linear, and costs are strictly convex. The exact sufficient conditions for the uniqueness of a solution are withheld, but for the solution not to be unique, similar to the case with Shannon Entropy studied by \citeA{matmck15}, a very rigid form of co-movement is required between payoffs and states.

\hypertarget{sec5}{ }

\section{Comparisons with the Standard Model}

In this section we compare and contrast the choice behavior that is produced by RI with Shannon Entropy and the choice behavior produced by the RI model developed in \hyperlink{sec4}{Section 4} that uses the MSSE measure developed in \hyperlink{sec3}{Section 3}. We first discuss the relationship between RU models and RI with MSSE, and then revisit the two motivating examples, \hyperlink{sec2.1}{Example 1} and \hyperlink{sec2.2}{Example 2}, from \hyperlink{sec2}{Section 2}.

\hypertarget{sec5.1}{ }
\subsection{Comparison with Random Utility Model}

It is standard practice to use a RU model to describe discrete choice settings. In such a model, the agent picks the option with the largest sum $u_n = v_n + \epsilon_n$ over all options $n\in\mathcal{N}$. Generally, $u_n$ represents the value of the option to the agent, $v_n$ represents the average value of the option across agents, and $\epsilon_n$ represents an idiosyncratic value to the agent. The role $\epsilon_n$ plays is up to interpretation, however, and is determined by the researchers specification \cite{tra09}. In a setting where agents are thought to be rationally inattentive, the above terms are interpreted in a different way because the agent's noisy behavior is generated by perceptual error instead of idiosyncratic differences in taste. In such settings, $u_n$ represents the perceived value to the agent, $v_n$ represents the true value to the agent, and $\epsilon_n$ is interpreted as an unobservable perceptual error that results from the noisy information strategy selected by the agent. \citeA{woo14} argues that this latter interpretation is necessary in many contexts due to the stochastic responses observed in perceptual discrimination tasks such as those administered by \citeA{deanel17}, which are akin to our \hyperlink{sec2.1}{Example 1} in \hyperlink{sec2.1}{Section 2.1}. While the interpretation of $\epsilon_n$ is relevant for welfare analysis, it is inconsequential for the description of choice behavior. How then can MSSE be interpreted in terms of an RU framework, and what insights may be provided about the fitting of RU models?

\citeA{matmck15} point out that choice probabilities predicted by RI with Shannon Entropy correspond to multinomial logit choice probabilities where it is as if option values have been shifted due to the agent's prior about potential values. An option that seems more desirable \textit{a priori} is more likely to be selected by the agent in every state of the world, and thus is overvalued by a multinomial logit regression.

Rational inattention with MSSE takes this one step further, as is shown by \hyperlink{th3}{Theorem 3}, allowing the shift in perceived value to also depend on easier to observe information sources (binary partitions associated multipliers that are less than $\lambda_M$). This flexibility seems natural in many real world environments. Consider an agent that is trying to select a restaurant to go to. One may expect that the chance of the agent selecting a given option to increase not only with the quality of the restaurant, and their prior impression of it, but also with easy to observe information such as on-line ratings the restaurant may have received.

\bigskip
\hypertarget{th3}{}
\noindent\textbf{Theorem 3:}

The choice behavior described by $\mathbb{P}$, a solution to (\hyperlink{7}{7}) subject to (\hyperlink{8}{8}) and (\hyperlink{9}{9}), is identical to the behavior produced by an RU model where each option $n\in\mathcal{N}$ has perceived value:
$$u_n  = \tilde{v}_n + \alpha_n + \epsilon_n,
$$
\noindent where $\tilde{v}_n=\dfrac{\textbf{v}_n(\omega)}{\lambda_M}$, $\epsilon_n$ has an iid Gumbel distribution, and:
$$\alpha_n= \mathsmaller{\mathsmaller{\frac{\lambda_1}{\lambda_M}}} \log (N \text{Pr}(n)) + \mathsmaller{\mathsmaller{\frac{\lambda_2-\lambda_1}{\lambda_M}}}\log( N \text{Pr}(n| \mathcal{P}_{\lambda_1}(\omega)))+\dots+ \mathsmaller{\mathsmaller{\frac{\lambda_M-\lambda_{M-1}}{\lambda_M}}}\log( N \text{Pr}(n|\cap_{i=1}^{M-1} \mathcal{P}_{\lambda_i}(\omega))).
$$

\hyperlink{th3}{Theorem 3} is meant to provide insight into the outcome of attempting to fit a RU model in an environment where agents are rationally inattentive with a cost function for information described by MSSE. \hyperlink{th3}{Theorem 3} does not say that a model of RI with MSSE is equivalent to a RU model. Even if choice data from a given choice problem cannot be used to reject one for the other, across choice problems MSSE produces behavior that can reject the hypothesis of a RU model. With MSSE, for instance, as with standard Shannon Entropy, adding an option can increase the chance of an existing option being selected, which is not possible with a RU model.

Also, it is worth mentioning that since optimal behavior may result in some options being selected with probability zero, \hyperlink{th3}{Theorem 3} implicitly defines each $\alpha_n$ on the extended reals so that $\alpha_n=-\infty$ if $\text{Pr}(n)=0$.\footnote{It can be shown that if optimal behavior results in $\text{Pr}(n)>0$, then $\text{Pr}(n|\omega)>0\,\,\forall\omega\in \Omega$. See \cite{wal19b}.}

\hypertarget{sec5.2}{ }

\subsection{Example 1 Revisited}

We now revisit \hyperlink{sec2.1}{Example 1} from \hyperlink{sec2.1}{Section 2.1}, which is described in \hyperlink{ta1}{Table 1}. It seems natural that it should be easier for the agent to answer the question `Are 60 of the balls blue?', than it is for them to answer `Are 51 or more of the balls blue?'. Similarly, it seems natural that it should be easier for the agent to answer the question `Are 60 of the balls red?', than it is for them to answer `Are 51 or more of the balls red?'. Symmetry also means that the questions `Are 60 of the balls blue?' and `Are 60 of the balls red?' should have the same expected cost, and the questions `Are 51 or more of the balls blue?' and `Are 51 or more of the balls red?' should have the same expected cost. We can thus assume $\mathcal{P}_{\lambda_1} = \{A_1,\,A_2,\,A_3\} = \{\{\omega_1\},\,\{\omega_2\cup\omega_3\},\,\{\omega_4\}\}$, and $\mathcal{P}_{\lambda_2}  = \{\{\omega_1\cup\omega_2\},\,\{\omega_3\cup\omega_4\}\}$.

Solutions to \hyperlink{co1}{Corollary 1} combined with \hyperlink{th2}{Theorem 2} mean that the chance of the agent selecting option 1 is increasing in the number of blue balls, as can be seen in \hyperlink{fi1}{Figure 1}, which depicts optimal behavior in each state of the world for a range of $\lambda_1$. When $\lambda_1$ is small relative to $\lambda_2$ the agent chooses option 1 in state $\omega_1$ with a high probability, and choose option 2 in state $\omega_4$ with a high probability. The agent is thus better able to discern the state of the world when there are 40 of one color ball and 60 of the other than when there are 49 of one color and 51 of the other. This is supported by the experimental work of \citeA{deanel17}, and is in contrast with the behavior predicted by a model of RI that uses Shannon Entropy. 

\citeA{moryan16} identify a related issue with Shannon Entropy's lack of perceptual distance, and warn against its use in some continuous settings because it predicts discontinuous changes in behavior at places where payoffs change discontinuously. In the limit, as the number of different perceptual distances is allowed to grow, MSSE can be used to produce the kind of continuous behavior that \citeA{moryan16} desire.

\begin{figure}[h]
\begin{center}
\hypertarget{fi1}{}
\caption{}
\includegraphics[scale=0.62]{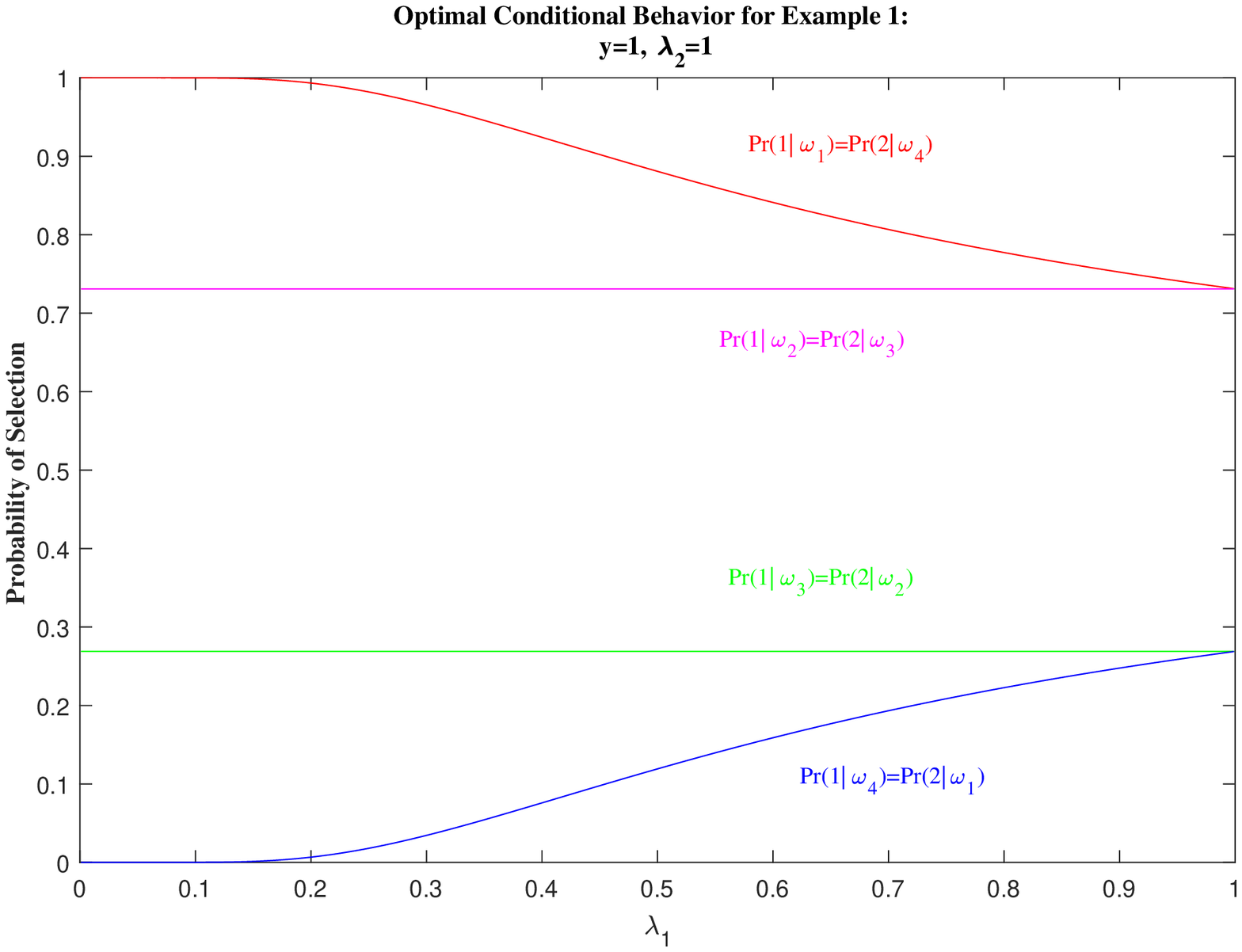}
\end{center}
\end{figure}

\hypertarget{sec5.3}{ }

\subsection{Example 2 Revisited}
We now revisit \hyperlink{sec2.2}{Example 2} from \hyperlink{sec2.2}{Section 2.2}, which is described in \hyperlink{ta2}{Table 2}. We assumed that learning the value of option 1 is less costly than learning the value of option 2. That is to say, answering the question `Is option 1 of value $H$?' has a lower expected cost to the agent than the question `Is option 2 of value $H$?'. We can thus assume: $\mathcal{P}_{\lambda_1} =\{A_1,\,A_2\} = \{\{\omega_1\cup\omega_2\},\,\{\omega_3\cup\omega_4\}\}$, and $\mathcal{P}_{\lambda_2} = \{\{\omega_1\cup\omega_3\},\,\{\omega_2\cup\omega_4\}\}$.

Solutions to \hyperlink{co1}{Corollary 1} in this environment for a range of $\lambda_1$ can be found in \hyperlink{fi2}{Figure 2}, which shows that when $\lambda_1$ is small compared to $\lambda_2$, the agent selects option 1 with a high probability when it is of value $H$, and selects option 2 with a high probability when option 1 is of value $L$. As $\lambda_1$ increases relative to $\lambda_2$, the chance of option 1 being selected when it is of value $H$ decreases. Similarly, as $\lambda_1$ increases relative to $\lambda_2$, the chance of option 1 being selected when it is of value $L$ increases. Note that the solutions to \hyperlink{co1}{Corollary 1} mean that the agent is more likely to select option 1 when state $\omega_1$ has been realized since $\text{Pr}(1|A_1)>\text{Pr}(2|A_1)$, and more likely to select option 2 when state $\omega_4$ has been realized since $\text{Pr}(1|A_2)<\text{Pr}(2|A_2)$, as can be observed with \hyperlink{th2}{Theorem 2}.

\begin{figure}
\begin{center}
\hypertarget{fi2}{}
\caption{}
\includegraphics[scale=0.62]{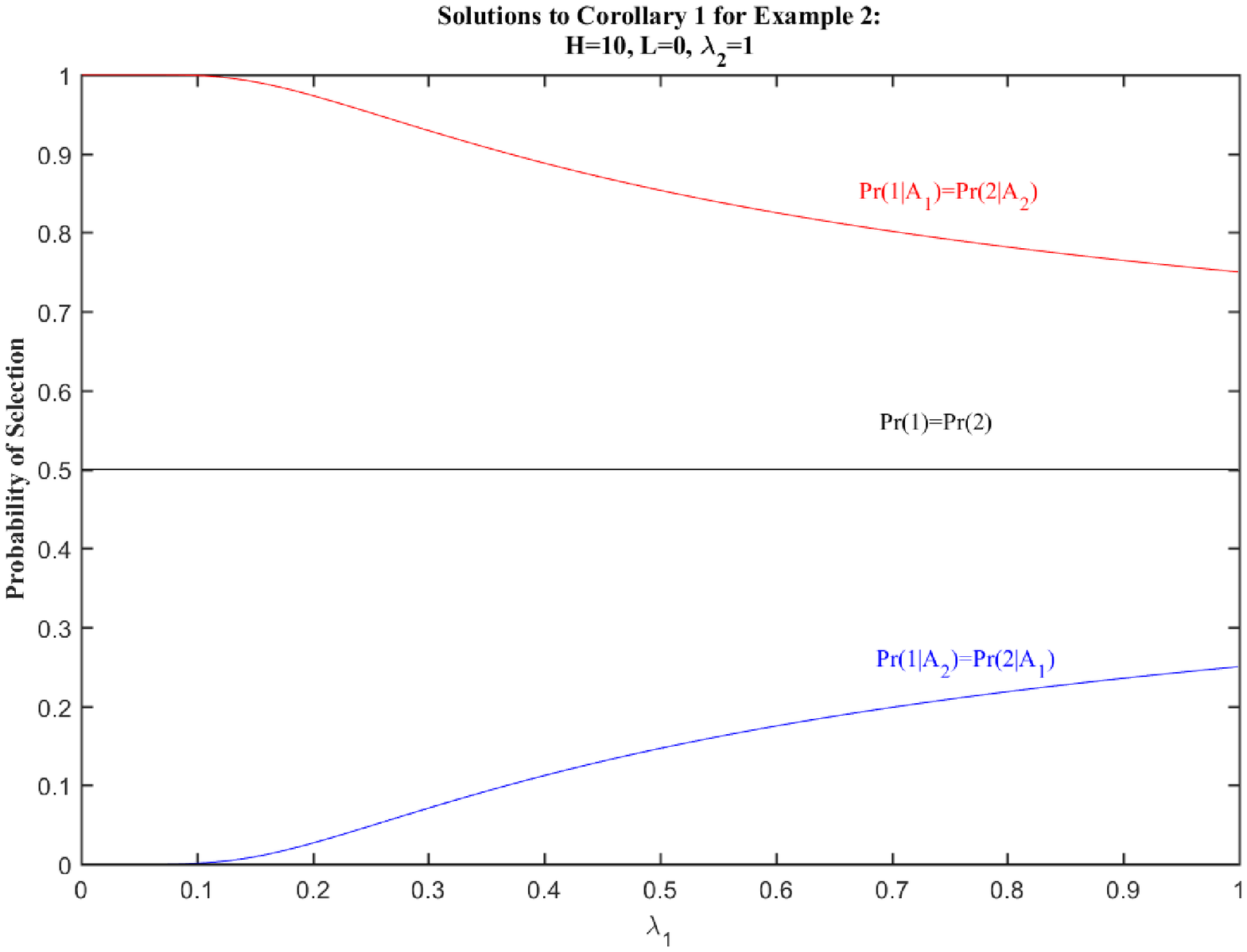}
\end{center}
\end{figure}

Solutions to \hyperlink{co1}{Corollary 1} combined with \hyperlink{th3}{Theorem 3} mean that if an econometrician tries to fit this environment with a multinomial logit model that their estimate of $H_1$, the high value of option 1, is biased upwards by $\frac{\lambda_2-\lambda_1}{\lambda_2} \log(2\text{Pr}(1|A_1))$, which is greater than zero since $\text{Pr}(1|A_1)>1/2$, and their estimate of $L_1$, the low value of option 1, is biased downwards by $\frac{\lambda_2-\lambda_1}{\lambda_2} \log(2\text{Pr}(1|A_2))$, which is less than zero since $\text{Pr}(1|A_2)<1/2$. These biases are despite the fact that the unconditional chance of either option being selected is the same: $\text{Pr}(1)=\text{Pr}(2)=1/2$. As such, the econometrician may have believed their analysis was not susceptible to informational biases if they had used Shannon Entropy to model the environment.

\hypertarget{sec6}{ }
\section{Conclusion}

Rational inattention models that use Shannon Entropy to measure the cost of learning demonstrate that informational biases in random utility models can be significant for welfare and counterfactual analysis. The biases that have previously been identified in the literature are independent of the realized state of the world, depending only on the agent's prior about the environment. These previously identified biases manifest themselves in the unconditional choice probabilities of the agent. 

This paper contributes to the literature by proposing and axiomatizing a new measure of uncertainty that features perceptual distance, maintains much of the tractability of Shannon's standard measure, and identifies a new kind of informational bias. The new form of bias can be present even when the agent has the same unconditional chance of selecting each option, which may seem to indicate an unbiased environment based on the previous literature.

\newpage
\hypertarget{ap1}{ }
\section*{Appendix 1}

Before we prove \hyperlink{le1}{Lemma 1}, we show some other useful results:

\bigskip

\hypertarget{le4}{ }
\noindent \textbf{Lemma 4.} If a partition $\tilde{\mathcal{P}}$ is coarser than a learning strategy invariant partition $\mathcal{P}$, then $\tilde{\mathcal{P}}$ is also learning strategy invariant.

\smallskip

\hypertarget{pl4}{ }
\noindent \textbf{Proof.} Suppose $\mathcal{P}$ is a learning strategy invariant partition, and $\tilde{\mathcal{P}}$ is coarser than ${\mathcal{P}}$. If $\tilde{\mathcal{P}}={\mathcal{P}}$ we are done. 

If $\tilde{\mathcal{P}}\neq {\mathcal{P}}$, then the definition of learning strategy invariance tells us that for any learning strategy $\tilde{S}=(\mathcal{P}_1,\,\dots,\,\mathcal{P}_n)$ such that $\sigma(\tilde{S})=\sigma(\tilde{P})$, and any $\mu$:
$$C({\mathcal{P}},\,\mu) =  C((\tilde{\mathcal{P}},\, \mathcal{P}),\,\mu) = C(\tilde{\mathcal{P}},\,\mu) + \mathbb{E}[C(\mathcal{P},\,\mu (\cdot |\tilde{\mathcal{P}}(\omega)))],
$$ 
\noindent and,
$$C({\mathcal{P}},\,\mu) = C(\tilde{S},\,\mu) + \mathbb{E}[C(\mathcal{P},\,\mu (\cdot |\mathlarger{\cap}_{i=1}^n\mathcal{P}_i(\omega)))]
 = C(\tilde{S},\,\mu) + \mathbb{E}[C(\mathcal{P},\,\mu (\cdot |\tilde{\mathcal{P}}(\omega)))].
$$
\noindent Thus, $ C(\tilde{\mathcal{P}},\,\mu)=C(\tilde{S},\,\mu)$ for all such $\tilde{S}$, and any $\mu$, so $\tilde{\mathcal{P}}$ is also learning strategy invariant.$\blacksquare$

\bigskip

\hypertarget{le5}{ }
\noindent \textbf{Lemma 5.} If $\mathcal{P}=\{A_1,\,\dots,\,A_{m}\}$ is a learning strategy invariant partition, probability measure $\mu$ assigns a probability of one to an event $A_i\in  \{A_1,\,\dots,\,A_{m}\}$, and $C$ satisfies \hyperlink{ax5}{Axiom 5}, then $C({\mathcal{P}},\,\mu)=0$.

\smallskip

\hypertarget{pl5}{ }
\noindent \textbf{Proof.} Suppose $\mathcal{P}=\{A_1,\,\dots,\,A_{m}\}$ is a learning strategy invariant partition of the state space $\Omega$, and there is an $A_i\in  \{A_1,\,\dots,\,A_{m}\}$ such that $\mu(A_i)=1$. It is without loss to further assume $i=1$. If $m>2$, or there is a learning strategy invariant partition $\tilde{\mathcal{P}}\neq \mathcal{P}$ of $\Omega$ which is finer than $\mathcal{P}$, then we can show \hyperlink{le5}{Lemma 5} without any axioms.

If $m>2$, let $\tilde{\mathcal{P}}=\{A_1, A_2,\,(A_1\cup A_2)^c\}$, $\hat{\mathcal{P}}=\{A_1\cup A_2,\,A_3,\,\dots ,\,A_m\}$, $S_1=(\tilde{\mathcal{P}},\,\hat{\mathcal{P}})$, and $S_2=(\tilde{\mathcal{P}},\,\hat{\mathcal{P}},\,\mathcal{P})$. The definition of learning strategy invariance tells us $C(S_1,\,\mu)=C(S_2,\,\mu)$, so $C({\mathcal{P}},\,\mu)=0$.

If $m=2$, and learning strategy invariant partition $\tilde{\mathcal{P}}\neq \mathcal{P}$ of $\Omega$ is finer than $\mathcal{P}$, then let $S_1=(\tilde{\mathcal{P}})$, and $S_2=(\mathcal{P},\,\tilde{\mathcal{P}})$. The definition of learning strategy invariance tells us $C(S_1,\,\mu)=C(S_2,\,\mu)$, so $C({\mathcal{P}},\,\mu)=0$.

If $m=2$, and there is not a learning strategy invariant $\tilde{\mathcal{P}}\neq \mathcal{P}$ of $\Omega$ which is finer than $\mathcal{P}$, then we can invoke \hyperlink{ax5}{Axiom 5}, and use a similar argument as the previous paragraph to get the desired result (if $C({\mathcal{P}},\,\mu)\neq 0$ when $\mu$ assigns a probability of one to an event in $\mathcal{P}$, then we are rejecting the potential of a finer learning strategy invariant partition, with three or more events in it, of some finer state space $\tilde{\Omega})$).$\blacksquare$

\bigskip

\noindent \textbf{Proof of \hyperlink{le1}{Lemma 1}.} Suppose $\mathcal{P}=\{A_1,\,\dots,\,A_{m}\}$ is a learning strategy invariant partition of the state space $\Omega$. If $m>2$, or there is a partition $\tilde{\mathcal{P}}\neq \mathcal{P}$ of $\Omega$ which is finer than $\mathcal{P}$, then we can show \hyperlink{le1}{Lemma 1} holds using only \hyperlink{ax4}{Axiom 4}. 

Suppose $m>2$. If we show that for any $i,\,j\in \{1,\,\dots,\,{m}\}$ such that $i\neq j$, $C({\mathcal{P}},\,\mu)=  C({\mathcal{P}},\,\tilde{\mu})$ if $\mu(A_k)=\tilde{\mu}(A_k)$ for $k\notin\{i,\,j\}$, $\mu(A_i)=\tilde{\mu}(A_j)$, and $\mu(A_j)=\tilde{\mu}(A_i)$, then the desired result holds, since a series of pairwise switches like this can be used to create any permutation desired. It is without loss to assume $i=1$ and $j=2$. Define $\tilde{\mathcal{P}}=\{A_1, A_2,\,(A_1\cup A_2)^c\}$. Notice that $\tilde{\mathcal{P}}$ must be learning strategy invariant based on  \hyperlink{le4}{Lemma 4}. Further, if we show that $C(\tilde{\mathcal{P}},\,\mu)=C(\tilde{\mathcal{P}},\,\tilde{\mu})$, then $C({\mathcal{P}},\,\mu)=C({\mathcal{P}},\,\tilde{\mu})$, since, if we define $\hat{\mathcal{P}}=\{A_1\cup A_2,\,A_3,\,\dots ,\,A_m\}$, then by \hyperlink{ax4}{Axiom 4}:
$$C({\mathcal{P}},\,\mu) = C(\tilde{\mathcal{P}},\,\mu) + \mathbb{E}[C(\hat{\mathcal{P}},\,\mu (\cdot |\tilde{\mathcal{P}}(\omega)))]
$$
$$= C(\tilde{\mathcal{P}},\,\tilde{\mu}) + \mathbb{E}[C(\hat{\mathcal{P}},\,\tilde{\mu} (\cdot |\tilde{\mathcal{P}}(\omega)))] = C({\mathcal{P}},\,\tilde{\mu}).
$$ 
\noindent Now, let $\mathcal{P}^b_1=\{A_1,A_1^c\}$, $\mathcal{P}^b_2= \{A_2,A_2^c\}$, and $\mathcal{P}^b_3 =\{A_1\cup A_2,\,(A_1\cup A_2)^c\}$. Notice $\mathcal{P}^b_1$, $\mathcal{P}^b_2$ and $\mathcal{P}^b_3$, are all coarser than $\tilde{\mathcal{P}}$. Then, since $\tilde{\mathcal{P}}$ is learning strategy invariant:
$$C(\tilde{\mathcal{P}},\,{\mu}) = C({\mathcal{P}^b_3},\,{\mu}) + \mathbb{E}[C({\mathcal{P}^b_1},\,{\mu}(\cdot |\mathcal{P}^b_1(\omega))],
$$
\noindent and,
$$C(\tilde{\mathcal{P}},\,\tilde{\mu}) = C({\mathcal{P}^b_3},\,\tilde{\mu}) + \mathbb{E}[C({\mathcal{P}^b_1},\,\tilde{\mu}(\cdot |\mathcal{P}^b_1(\omega))].
$$
Notice that \hyperlink{ax4}{Axiom 4} tells us $C({\mathcal{P}^b_3},\,{\mu})=C({\mathcal{P}^b_3},\,\tilde{\mu})$. So, all that remains to show is that if the probability measure $\tilde{\nu}$ is a permutation of the probability measure ${\nu}$ on $\mathcal{P}^b_1$, then $C({\mathcal{P}}^b_1,\,{\nu})=C({\mathcal{P}}^b_1,\,\tilde{\nu})$. Fix arbitrary $\nu(A_1)=x\in [0,\,1]$. Now consider the probability measures $q_1,\,q_2,\,q_3$, such that: $$q_1(A_1)=x,\,\,q_1(A_2)=0,\,\,q_1((A_1\cup A_2)^c)=1-x,$$
$$q_2(A_1)=0,\,\,q_2(A_2)=x,\,\,q_2((A_1\cup A_2)^c)=1-x,$$ 
$$q_3(A_1)=1-x,\,\,q_3(A_2)=x,\,\,q_3((A_1\cup A_2)^c)=0.$$
\noindent Notice that $q_3$ is a permutation of $q_1$ on $ \mathcal{P}^b_1$. So then, using \hyperlink{ax4}{Axiom 4}, the definition of learning strategy invariance, and \hyperlink{le5}{Lemma 5}, all repeatedly:
$$C({\mathcal{P}}^b_1, q_1)=C(\tilde{\mathcal{P}}, q_1) = C({\mathcal{P}}^b_3, q_1)=C({\mathcal{P}}^b_3, q_2)
$$
$$=C(\tilde{\mathcal{P}}, q_2)=C({\mathcal{P}}^b_2, q_2)=C({\mathcal{P}}^b_2, q_3)=C(\tilde{\mathcal{P}}, q_3)=C({\mathcal{P}}^b_1, q_3),
$$ 
\noindent and we are done.

If $m=2$, and there is a learning strategy invariant partition $\tilde{\mathcal{P}}\neq \mathcal{P}$ of $\Omega$ which is finer than $\mathcal{P}$, then the first half of the proof establishes permutations of the probability measure $\mu$ on $\tilde{\mathcal{P}}$ do not change $C(\tilde{\mathcal{P}},\,\mu)$. Denote $\mathcal{P}=\{A_1,\,A_2\}$. By definition of a finer partition, there is $A_i,A_j\in \tilde{\mathcal{P}}$ such that $A_i\subseteq A_1$ and $A_j\subseteq A_2$. Define $\mu$ so that $\mu(A_i) + \mu(A_j)=1$, and probability measure $\tilde{\mu}$ so that $\tilde{\mu}(A_i) = \mu(A_j)$ and $\tilde{\mu}(A_j) = \mu(A_i)$. Then, learning strategy invariance and \hyperlink{le5}{Lemma 5} tell us:
$$C({\mathcal{P}},\,\mu)=C(\tilde{\mathcal{P}},\,\mu)=C(\tilde{\mathcal{P}},\,\tilde{\mu})=C({\mathcal{P}},\,\tilde{\mu}).
$$
\noindent Thus, since this works for any such $\mu$, \hyperlink{ax4}{Axiom 4} tells us permutations of $\mu$ on $\mathcal{P}$ do not change $C({\mathcal{P}},\,\mu)$.

If $m=2$, and there is not a learning strategy invariant partition $\tilde{\mathcal{P}}\neq \mathcal{P}$ of $\Omega$ which is finer than $\mathcal{P}$, then we can invoke \hyperlink{ax5}{Axiom 5} and use a similar argument as the previous paragraph to get the desired result (if $C({\mathcal{P}},\,\mu)$ is not constant with respect to permutations of $\mu$ on $\mathcal{P}$, then we are rejecting the potential of a finer learning strategy invariant partition, with three or more events in it, of some finer state space $\tilde{\Omega})$).$\blacksquare$

\bigskip

\hypertarget{pl2}{ }
\noindent \textbf{Proof of \hyperlink{le2}{Lemma 2}.}

For all partitions $\mathcal{P}=\{A_1,\,\dots,\,A_m\}$ and probability measures $\mu$ defined on $\mathcal{P}$, define the reverse order statistic $\mu(\mathcal{P})=(\mu_{(1)}(\mathcal{P}),\,\dots,\,\mu_{(m)}(\mathcal{P}))$ to be the unique vector of probabilities such that $\mu_{(1)}(\mathcal{P})\geq \,\dots\,\geq \mu_{(m)}(\mathcal{P})$, and if we define probability measure $\tilde{\mu}$ so that $\tilde{\mu}(A_i)=\mu_{(i)}(\mathcal{P})\,\,\forall i\in\{1,\,\dots,\,A_m\}$, then $\tilde{\mu}$ is a permutation of $\mu$ on $\mathcal{P}$.

Suppose $\mathcal{P}_i=\{A_1,\,\dots,\,A_{m}\}$ is a learning strategy invariant partition, and $\mathcal{P}_j=\{A_1^j,\,\dots,\,A_{m_j}^j\}$ is another learning strategy invariant partition such that $\tilde{\mathcal{P}}_i\neq \mathcal{P}_i$, and $\tilde{\mathcal{P}}_i$ is either finer or coarser than $\mathcal{P}_i$. \hyperlink{le1}{Lemma 1} and \hyperlink{ax4}{Axiom 4} tells us that $C({\mathcal{P}},\,\mu)$ and $C(\tilde{\mathcal{P}},\,\mu)$ are fully determined by $\mu(\mathcal{P})$ and $\mu(\tilde{\mathcal{P}})$ respectively, and if the strictly positive entries of $\mu(\mathcal{P})$ and $\mu(\tilde{\mathcal{P}})$ are the same, then $C({\mathcal{P}},\,\mu)=C(\tilde{\mathcal{P}},\,\mu)$. What does this mean? This means that there is a function which maps from vectors of probabilities onto the reals, $c_i:\cup_{j=i}^{\infty} \triangle^j \rightarrow \mathbb{R},$ where $\triangle^j$ is the $j$ simplex, such that for any learning strategy invariant partition $\tilde{\mathcal{P}}_i$, such that $\tilde{\mathcal{P}}_i$ is either finer or coarser than $\mathcal{P}_i$, then $C(\tilde{\mathcal{P}},\,\mu) = c_i(\mu(\tilde{\mathcal{P}}))$. We define $c_i$ on $\triangle^j$ for arbitrarily large $j$ because if there is a partition $\hat{\mathcal{P}}_i$ of some finer state space $\hat{\Omega}$, that is finer than $\mathcal{P}$, that is composed of $j$ events, then $c_i$ is defined on $\triangle^j$, and if we assume $c_i$ is not defined on $\triangle^j$, we contradict \hyperlink{ax5}{Axiom 5}.

The rest of the proof follows the work of \citeA{sha48} closely. Define $h$ so for $n\in\mathbb{N}$, $h(n)\equiv c_i(1/n,\,\dots,\,1/n,0)$. We do this for all  $n\in\mathbb{N}$ so we do not violate \hyperlink{ax5}{Axiom 5}. Learning strategy invariance implies $h(s^r) = r\cdot h(s)$, which is reminiscent of logarithms, and is some nice foreshadowing for the rest of the proof. Given arbitrarily small $\epsilon>0$, and integers $s$ and $t$, pick $n$ and $r$ so that $2/n<\epsilon$, and $s^r \leq t^n< s^{r+1}$. So:
$$r\log (s) \leq n \log (t) < (r+1)\log (s)\Longrightarrow\dfrac{r}{n}\leq \dfrac{\log (t)}{\log (s)} < \dfrac{r+1}{n} \Longrightarrow \mathlarger{\mathlarger{\mathlarger{|}}}\dfrac{r}{n} - \dfrac{\log (t)}{\log (s)}\mathlarger{\mathlarger{\mathlarger{|}}} < \dfrac{1}{n}.
$$
\noindent \hyperlink{ax2}{Axiom 2} then tells us:
$$h(s^r)\leq h(t^n) < h(s^{r+1})\Longrightarrow r\cdot h(s)\leq n\cdot h(t) < (r+1)h(s)
$$
$$ \Longrightarrow \dfrac{r}{n} \leq \dfrac{h(t)}{h(s)}< \dfrac{r+1}{n} \Longrightarrow  \mathlarger{\mathlarger{\mathlarger{|}}}\dfrac{r}{n} - \dfrac{h (t)}{h (s)}\mathlarger{\mathlarger{\mathlarger{|}}} < \dfrac{1}{n}.
$$
\noindent All of this tells us:
$$\mathlarger{\mathlarger{\mathlarger{|}}}\dfrac{h(t)}{h(s)} - \dfrac{\log (t)}{\log (s)}\mathlarger{\mathlarger{\mathlarger{|}}} < \epsilon,$$
\noindent which can be shown to be true $\forall \epsilon>0$, and thus $h(n)= \lambda_i \log (n)$, where $\lambda_i$ must be a positive constant to satisfy \hyperlink{ax2}{Axiom 2}.

Let $p_k=\mu(A_k)$ for each $A_k\in\mathcal{P}_i$. Suppose, for now, that each $p_k$ is a rational number. Then there exists integers $n_1,\,\dots,\,n_m$, such that for all $k\in\{1,\,\dots,\,m\}$ we have:
$$p_k = \dfrac{n_k}{\sum\limits_{j=1}^m n_j}.
$$
\noindent Our interpretation is that we have a uniform distribution over $\sum\limits_{j}n_j$ equally likely states, and the chance of the event which happens with probability $p_k$ is the chance of one of the $n_k$ associated states occurring. Then using the definition of learning strategy invariance:
$$c_i\Bigg(\dfrac{1}{\sum\limits_{j}n_j},\,\dots,\,\dfrac{1}{\sum\limits_{j}n_j}\Bigg) = h\Bigg(\sum\limits_{j=1}^{m} n_j\Bigg) = \lambda_i\log\Bigg(\sum\limits_{j=1}^{m}n_j\Bigg) = c_i(p_1,\,\dots,\,p_m) + \sum\limits_{j=1}^{m}p_j \lambda_i\log(n_j),
$$
$$\Longrightarrow c_i(p_1,\,\dots,\,p_m) = \lambda_i\log\Bigg(\sum\limits_{j=1}^m n_j\Bigg) - \sum\limits_{j=1}^{m}p_j \lambda_i \log(n_j) 
$$
$$= \sum\limits_{k=1}^{m}\mathlarger{\mathlarger{\Bigg(}} p_k\lambda_i\log\Bigg(\sum\limits_{j=1}^{m}n_j\Bigg)\mathlarger{\mathlarger{\Bigg)}} - \sum\limits_{j=1}^{m}p_j\lambda_i \log(n_j) 
$$
$$=- \sum\limits_{k=1}^{m}p_k\lambda_i \log\Bigg( \dfrac{n_k}{\sum\limits_j n_j}\Bigg) =  - \lambda_i \sum\limits_{k=1}^{m}p_k \log (p_k) = \lambda_i \mathcal{H}(\mathcal{P}_i,\,\mu),
$$
\noindent where $\mathcal{H}$ is defined as in equation (\hyperlink{1}{1}). If any of the $p_i$ are irrational, then the density of the rationals and \hyperlink{ax3}{Axiom 3} can be used to get the same result. Thus:
$$ C(\mathcal{P}_i,\,\mu) = c_i(\mu (A_1),\,\dots,\,\mu (A_{m})) = \lambda_i \mathcal{H}(\mathcal{P}_i,\,\mu). \blacksquare
$$

\subsection*{Mutual Information}

Consider two partitions $\mathcal{P}_1$ and $\mathcal{P}_2$. Given some probability measure $\mu$, define the \textbf{mutual information} between $\mathcal{P}_1$ and $\mathcal{P}_2$, denoted $I(\mathcal{P}_1,\,\mathcal{P}_2,\,\mu)$, to be:
$$I(\mathcal{P}_1,\,\mathcal{P}_2,\,\mu) = \sum\limits_{a_1\in \mathcal{P}_1} \sum\limits_{a_2\in \mathcal{P}_2} \mu(a_1\cap a_2)\log \Big(\dfrac{\mu(a_1\cap a_2)}{\mu(a_1)\mu(a_2)}   \Big)
$$
\noindent Then, as is well known in the literature:
$$\mathcal{H}(\times\{\mathcal{P}_i\}_{i=1}^2,\,\mu)= \mathcal{H}(\mathcal{P}_1,\,\mu) + \mathcal{H}(\mathcal{P}_2,\,\mu) -I(\mathcal{P}_1,\,\mathcal{P}_2,\,\mu)$$
$$ = \equalto{\mathbb{E}[\mathcal{H}(\mathcal{P}_1,\,\mu(\cdot|\mathcal{P}_{2}(\omega)))]}{\mathcal{H}(\mathcal{P}_1,\,\mu) - I(\mathcal{P}_1,\,\mathcal{P}_2,\,\mu)} + I(\mathcal{P}_1,\,\mathcal{P}_2,\,\mu) + \equalto{\mathbb{E}[\mathcal{H}(\mathcal{P}_2,\,\mu(\cdot|\mathcal{P}_{1}(\omega)))]}{\mathcal{H}(\mathcal{P}_2,\,\mu)- I(\mathcal{P}_1,\,\mathcal{P}_2,\,\mu)}$$
$$ = \mathcal{H}(\mathcal{P}_1,\,\mu) + \mathbb{E}[\mathcal{H}(\mathcal{P}_2,\,\mu(\cdot|\mathcal{P}_{1}(\omega)))] = \mathcal{H}(\mathcal{P}_2,\,\mu) + \mathbb{E}[\mathcal{H}(\mathcal{P}_1,\,\mu(\cdot|\mathcal{P}_{2}(\omega)))]$$

\noindent and note that the strict concavity of $\mathcal{H}$ means that $I(\mathcal{P}_1,\,\mathcal{P}_2,\,\mu)\geq0$. 

Mutual information can be thought of as the information that is double counted if one were to compute the total uncertainty about the outcome of $\mathcal{P}_1$ and $\mathcal{P}_2$ by simply adding up the uncertainty about the outcome of $\mathcal{P}_1$ and the uncertainty about the outcome of $\mathcal{P}_2$. When the mutual information increases and the individual uncertainty about the outcome of $\mathcal{P}_1$ and the outcome of $\mathcal{P}_2$ are held constant the total uncertainty about the outcome of $\mathcal{P}_1$ and $\mathcal{P}_2$ decreases because the amount that remains to be learned after observing one of the outcomes of either $\mathcal{P}_1$ or $\mathcal{P}_2$ decreases.

Mutual information can be acquired by learning the value of either $\mathcal{P}_1$ or $\mathcal{P}_2$. When we think of an agent that is trying to acquire information in an efficient fashion, we should always envision them acquiring mutual information from the cheapest source, by learning about whichever of $\mathcal{P}_1$ and $\mathcal{P}_2$ has the lowest associated multiplier. This logic is formalized by the result in \hyperlink{le6}{Lemma 6}.


%
%

\bigskip

\hypertarget{le6}{ }
\noindent \textbf{Lemma 6.} If $C$ satisfies our five axioms, and $S^b=\{\mathcal{P}_1^b,\,\dots,\,\mathcal{P}_i^b,\,\mathcal{P}_{i+1}^b,\,\dots,\,\mathcal{P}_m^b\}$ and $\tilde{S}^b=\{\mathcal{P}_1^b,\,\dots,\,\mathcal{P}_{i+1}^b,\,\mathcal{P}_{i}^b,\,\dots,\,\mathcal{P}_m^b\}$ are two binary learning strategies such that $\mathcal{P}_i^b$ and $\mathcal{P}_{i+1}^b$'s associated multipliers are ordered $\lambda_i\geq \lambda_{i+1} $, then for all probability measures $\mu$:
$$C(S^b,\,\mu) \geq C(\tilde{S}^b,\,\mu).
$$

\hypertarget{pl6}{ }
\noindent \textbf{Proof.} For all realizations of $\cap_{j=1}^{i-1}\mathcal{P}_j^b(\omega)$:
$$C((\mathcal{P}_i^b,\,\mathcal{P}_{i+1}^b),\,\mu(\cdot|\cap_{j=1}^{i-1}\mathcal{P}_j^b(\omega))) = \lambda_i\mathcal{H}(\mathcal{P}_i^b,\,\mu(\cdot|\cap_{j=1}^{i-1}\mathcal{P}_j^b(\omega))) + \lambda_{i+1}\mathbb{E}[\mathcal{H}(\mathcal{P}_{i+1}^b,\,\mu(\cdot|\cap_{j=1}^{i}\mathcal{P}_j^b(\omega)))]
$$
$$= \lambda_i\mathcal{H}(\mathcal{P}_i^b,\,\mu(\cdot|\cap_{j=1}^{i-1}\mathcal{P}_j^b(\omega))) + \lambda_{i+1}\Big(\mathcal{H}(\mathcal{P}_{i+1}^b,\,\mu(\cdot|\cap_{j=1}^{i-1}\mathcal{P}_j^b(\omega))) -I(\mathcal{P}_i^b,\,\mathcal{P}_{i+1}^b,\,\mu(\cdot|\cap_{j=1}^{i-1}\mathcal{P}_j^b(\omega))\Big)
$$
$$\geq \lambda_i\Big(\mathcal{H}(\mathcal{P}_i^b,\,\mu(\cdot|\cap_{j=1}^{i-1}\mathcal{P}_j^b(\omega)))-I(\mathcal{P}_i^b,\,\mathcal{P}_{i+1}^b,\,\mu(\cdot|\cap_{j=1}^{i-1}\mathcal{P}_j^b(\omega))\Big) + \lambda_{i+1}\mathcal{H}(\mathcal{P}_{i+1}^b,\,\mu(\cdot|\cap_{j=1}^{i-1}\mathcal{P}_j^b(\omega))) 
$$
$$=\lambda_{i+1}\mathcal{H}(\mathcal{P}_{i+1}^b,\,\mu(\cdot|\cap_{j=1}^{i-1}\mathcal{P}_j^b(\omega))) + \lambda_i\mathbb{E}[\mathcal{H}(\mathcal{P}_i^b,\,\mu(\cdot|(\cap_{j=1}^{i-1}\mathcal{P}_j^b(\omega))\cap \mathcal{P}_{i+1}^b(\omega)))] 
$$
$$= C((\mathcal{P}_{i+1}^b,\,\mathcal{P}_{i}^b),\,\mu(\cdot|\cap_{j=1}^{i-1}\mathcal{P}_j^b(\omega))).
$$

It is thus always weakly cheaper in expectation to have $\mathcal{P}_{i+1}$ before $\mathcal{P}_{i}$ since switching their order does not change the expected cost of implementing the binary partitions before or after the pair.$\blacksquare$

\bigskip

\hypertarget{pt1}{ }
\noindent \textbf{Proof of \hyperlink{th1}{Theorem 1}.} Given some probability measure $\mu$, suppose $S^b$ is a binary learning strategy such that $\sigma(S^b) = \mathcal{F}$, and $$C(S^b,\,\mu) = \min\limits_{S^b\in S^b(\Omega)}C(S^b,\,\mu).$$ We know such binary learning strategy exists whenever $C$ satisfies \hyperlink{ax1}{Axiom 1}. We may assume that if $\mathcal{P}_i^b$ and $\mathcal{P}_{i+1}^b$ are in $S^b$ with associated multipliers $\lambda_i$ and $\lambda_{i+1}$, that $\lambda_i\leq\lambda_{i+1}$. If not, then their order can be reversed and the resultant strategy is weakly less costly, as is shown in \hyperlink{le6}{Lemma 6}.

If for any $j\in\{1,\,\dots,\,M\}$, multiplier $\lambda_j$'s associated binary partitions $\mathcal{P}_i^b,\,\dots,\mathcal{P}_{i+k}^b$ in $S^b$ are such that $\sigma(\mathcal{P}_i^b,\,\dots,\mathcal{P}_{i+k}^b)\neq\sigma(\mathcal{P}_{\lambda_j}^b)$, then there are binary partitions $\mathcal{P}_{m+1}^b,\,\dots,\,\mathcal{P}_{m+l}^b$ with associated multiplier $\lambda_j$, such that $\sigma(\mathcal{P}_i^b,\,\dots,\mathcal{P}_{i+k}^b,\,\mathcal{P}_{m+1},\,\dots,\,\mathcal{P}_{m+l}^b)=\sigma(\mathcal{P}_{\lambda_j}^b)$. $\mathcal{P}_{m+1}^b,\,\dots,\,\mathcal{P}_{m+l}^b$ can be appended to the end of $S^b$, and the resultant strategy $\tilde{S}^b$ is also such that:
$$C(\tilde{S}^b,\,\mu) = \min\limits_{S^b\in S^b(\Omega)}C(S,\,\mu).$$ 
\noindent This is true since each appended binary partition has an expected cost of zero, since $\sigma(S^b) = \mathcal{F}$. \hyperlink{le6}{Lemma 6} then implies that if we reorder $\tilde{S}^b$ so that the new learning strategy $\hat{S}$'s binary partitions are ordered by their multipliers, then: 
$$C(\hat{S}^b,\,\mu)  = \min\limits_{S^b\in S^b(\Omega )}C(S,\,\mu).$$
\noindent We can thus assume that $S^b$ is such that for any $j\in\{1,\,\dots,\,M\}$ multiplier $\lambda_j$'s associated binary partitions $\mathcal{P}_i^b,\,\dots,\mathcal{P}_{i+k}^b$ in $S^b$ are such that $\sigma(\mathcal{P}_i^b,\,\dots,\mathcal{P}_{i+k}^b)=\sigma(\mathcal{P}_{\lambda_j})$.

For each $j\in\{1,\,\dots,\,M\}$ we thus have that if all binary partitions $\mathcal{P}_i^b,\,\dots,\mathcal{P}_{i+k}^b$ in $S^b$ with multiplier $\lambda_j$ are taken together that:
$$\mathbb{E}[C((\mathcal{P}_i^b,\,\dots,\mathcal{P}_{i+k}^b),\,\mu(\cdot|\cap_{t=1}^{i-1}\mathcal{P}_t^b(\omega)))] = \mathbb{E}\Big[\sum_{l=i}^{i+k} \lambda_j \mathcal{H}(\mathcal{P}_l^b,\,\mu(\cdot|\cap_{t=1}^{l-1}\mathcal{P}_t^b(\omega)))\Big]$$
$$ = \mathbb{E}[\lambda_j\mathcal{H}(\mathcal{P}_{\lambda_j},\,\mu(\cdot|\cap_{t=1}^{i-1}\mathcal{P}_t^b(\omega)))]=  \mathbb{E}[\lambda_j\mathcal{H}(\mathcal{P}_{\lambda_j},\,\mu(\cdot|\cap_{t=1}^{j-1}\mathcal{P}_{\lambda_t}(\omega)))].
$$
\noindent Where the second equality holds due to the properties of $\mathcal{H}$. This procedure can be carried out for all $\mu$. Thus:
$$ C(S^b,\,\mu) =  \min\limits_{S^b\in S^b(\Omega)}C(S,\,\mu).
$$
$$= \lambda_\mathsmaller{1} \mathcal{H}\Big(\mathcal{P}_{\lambda_1},\,\mu\Big)+ \mathlarger{\mathbb{E}}\bigg[\lambda_\mathsmaller{2}\mathcal{H}\Big(\mathcal{P}_{\lambda_2},\,\mu(\cdot|\mathcal{P}_{\lambda_1}(\omega))\Big)+\dots+\lambda_\mathsmaller{M} \mathcal{H}\Big(\mathcal{P}_{\lambda_M},\, \mu(\cdot|\mathlarger{\cap}_{{i=1}}^{{M-1}}\mathcal{P}_{\lambda_i}(\omega))\Big)\bigg]. \blacksquare
$$

\hypertarget{ap2}{ }

\section*{Appendix 2}

\hypertarget{pl3}{ }
 
\noindent \textbf{Proof of \hyperlink{le3}{Lemma 3}.} In \hyperlink{le3}{Lemma 3}, we show that we can rewrite the agent's problem in terms of selecting the choice probabilities described in equations (\hyperlink{4}{4}), (\hyperlink{5}{5}), and (\hyperlink{6}{6}). To do this, we first establish several other lemmas. In \hyperlink{le7}{Lemma 7}, we show that:$ \min\limits_{S\in S(\Omega)}C(S,\,\mu)$ is a strictly concave function of $\mu$. This is a commonly known property of Shannon Entropy, but needs to be established for in our context. This implies that $\textbf{C}$ is strictly convex. We then show, in \hyperlink{le8}{Lemma 8}, that, given the convexity of $\textbf{C}$, any selected action is associated with a particular posterior probability. This is desirable because it allows us to reduce the strategies considered to recommendation strategies. That is, we are able to focus on signals that are simply a recommendation of an option. In \hyperlink{le9}{Lemma 9}, we show that we may rewrite the cost function in terms of the choice probabilities in equations (\hyperlink{4}{4}), (\hyperlink{5}{5}), and (\hyperlink{6}{6}).

\bigskip
\hypertarget{le7}{}
\noindent \textbf{Lemma 7.} $\min\limits_{S\in S(\Omega)}C(S,\,\mu)$ is a strictly concave function of $\mu$. Namely, if there are probability measures $\mu_a$, and $\mu_b$, such that $\mu = \alpha \mu_a  + (1-\alpha)\mu_b$ for some $\alpha \in (0,1)$, and $\mu_a\neq\mu_b$, then: $$\min\limits_{S\in S(\Omega)}C(S,\,\mu) > \alpha \Big( \min\limits_{S\in S(\Omega)}C(S,\,\mu_a)\Big) + (1-\alpha)\Big(\min\limits_{S\in S(\Omega)}C(S,\,\mu_b)\Big).$$

\noindent \textbf{Proof.} For each probability measure $\mu$, $i\in\{1,\,\dots,\,M\}$, and realization of $\cap_{j=1}^{i-1}\mathcal{P}_{\lambda_j}(\omega)$, the strict concavity of Shannon Entropy \cite{matmck15, capea17} implies:
$$\mathcal{H}(\mathcal{P}_{\lambda_i}, \,\mu(\cdot|\cap_{j=1}^{i-1}\mathcal{P}_{\lambda_j}(\omega))) \geq \alpha\mathcal{H}(\mathcal{P}_{\lambda_i}, \,\mu_a(\cdot|\cap_{j=1}^{i-1}\mathcal{P}_{\lambda_j}(\omega))) + (1-\alpha)\mathcal{H}(\mathcal{P}_{\lambda_i}, \,\mu_b(\cdot|\cap_{j=1}^{i-1}\mathcal{P}_{\lambda_j}(\omega))).
$$
\noindent The inequality is also strict for at least one $i\in\{1,\,\dots,\,M\}$ since $\mu_a\neq\mu_b$. The desired result thus follows from \hyperlink{th1}{Theorem 1}.$\blacksquare$

\bigskip
\hypertarget{le8}{}
\noindent \textbf{Lemma 8.} If action $n \in \mathcal{N}$ is selected with positive probability, $\text{Pr}(n)>0$, as the outcome of information strategy $F$ with is a solution to (\hyperlink{2}{2}) subject to (\hyperlink{3}{3}), then there exists a posterior belief $B_n$ such that $F(\omega|s) = B_n$ with probability one whenever $n$ is selected.

\noindent \textbf{Proof.} It is impossible that there are two distinct sets of signals $S_n^1$ and $S_n^2$ which are observed with strictly positive probability, both of which lead to the selection of $n$, and induce different posteriors $F(\omega|s_1) \neq F(\omega|s_2)$ for $s_1\in S_n^1$ and $s_2\in S_n^2$. $\min\limits_{S\in S(\Omega)}C(S,\,\mu)$ is strictly concave in $\mu$, as shown in \hyperlink{le7}{Lemma 7}, so the agent could thus do better by replacing their original information strategy $F$ with a new information strategy $\tilde{F}$ which is identical to $F$ except the signals in $S_n^1$ and $S_n^2$ are replaced by $s_0$: $\forall \omega\in \Omega$ let $\tilde{F}(s_0|\omega) = \int\limits_{s\in S_n^1} F(s|\omega) + \int\limits_{s\in S_n^2}F(s|\omega)$. This is true because payoffs are linear, and the law of iterated expectations implies the agent still picks $n$ after $s_0$ is realized since $\forall \, \nu \in \mathcal{N}$:
$$\mathbb{E}_{\tilde{F}}[\mathbf{v}_n(\omega)|s_0]=  \dfrac{\sum\limits_{\omega\in\Omega}\int\limits_{s\in S_n^1} F(s|\omega)\mu(\omega)}{\sum\limits_{\omega\in\Omega}\bigg(\int\limits_{s\in S_n^1} F(s|\omega) \mu(\omega) + \int\limits_{s\in S_n^2} F(s|\omega)\mu(\omega)\bigg)} \mathbb{E}_F[\mathbf{v}_n(\omega)|s\in S_n^1]
$$
 $$+ \dfrac{\sum\limits_{\omega\in\Omega}\int\limits_{s\in S_n^2} F(s|\omega)\mu(\omega)}{\sum\limits_{\omega\in\Omega}\bigg(\int\limits_{s\in S_n^2} F(s|\omega)\mu(\omega) + \int\limits_{s\in S_n^2} F(s|\omega)\mu(\omega)\bigg)} \mathbb{E}_F[\mathbf{v}_n(\omega)|s\in S_n^2]$$
 $$\geq \dfrac{\sum\limits_{\omega\in\Omega}\int\limits_{s\in S_n^1} F(s|\omega)\mu(\omega)}{\sum\limits_{\omega\in\Omega}\bigg(\int\limits_{s\in S_n^1} F(s|\omega)\mu(\omega) + \int\limits_{s\in S_n^2} F(s|\omega)\mu(\omega)\bigg)} \mathbb{E}_F[\mathbf{v}_\nu(\omega)|s\in S_n^1]$$
$$+ \dfrac{\sum\limits_{\omega\in\Omega}\int\limits_{s\in S_n^2} F(s|\omega)\mu(\omega)}{\sum\limits_{\omega\in\Omega}\bigg(\int\limits_{s\in S_n^2} F(s|\omega)\mu(\omega) +\int\limits_{s\in S_n^2} F(s|\omega)\mu(\omega)\bigg)} \mathbb{E}_F[\mathbf{v}_\nu(\omega)|s\in S_2]= \mathbb{E}_{\tilde{F}}[\mathbf{v}_{\nu}(\omega)|s_0].\blacksquare
$$.

\bigskip
\hypertarget{le9}{}
\noindent \textbf{Lemma 9.} The cost of information for a given strategy in equation (\hyperlink{2}{2}) can be written:
$$\textbf{C}(F(s,\,\omega),\,\mu)=\textbf{C}(\mathbb{P},\, \mu) 
$$
$$=\sum\limits_{\omega\in\Omega} \mu(\omega)\sum\limits_{n\in \mathcal{N}} \Big( - \lambda_1 \text{Pr}(n) \log (\text{Pr}(n)) - (\lambda_2 - \lambda_1) \text{Pr}(n|\mathcal{P}_{\lambda_1}(\omega))\log(\text{Pr}(n|\mathcal{P}_{\lambda_1}(\omega)))$$
$$ - (\lambda_3 - \lambda_2) \text{Pr}(n|\mathcal{P}_{\lambda_1}(\omega)\cap\mathcal{P}_{\lambda_2}(\omega))\log(\text{Pr}(n|\mathcal{P}_{\lambda_1}(\omega)\cap \mathcal{P}_{\lambda_2}(\omega))) 
$$
$$-\,\dots\,- (\lambda_M - \lambda_{M-1})\text{Pr}(n|\cap_{i=1}^{M-1} \mathcal{P}_{\lambda_i}(\omega))\log(\text{Pr}(n|\cap_{i=1}^{M-1} \mathcal{P}_{\lambda_i}(\omega)))
+ \lambda_M \text{Pr}(n|\omega)\log(\text{Pr}(n|\omega))\Big).
$$

\hypertarget{pl7}{ }
\noindent \textbf{Proof.} Let $\mathcal{P}_s = (S_1,\,\dots,\,S_n)$ denote a partition of the space of signals the agent may receive. We showed in \hyperlink{le8}{Lemma 8} that for each $S_i$ if $s$ in $S_i$ then with probability one $s$ results in a particular posterior. We then have:
$$\textbf{C}(F(s,\,\omega),\,\mu) = \mathbb{E}[ \min\limits_{S\in S(\Omega)}C(S,\,\mu) - \min\limits_{S\in S(\Omega)}C(S,\,\mu(\cdot|s))]
$$
\hypertarget{11}{}
\begin{equation}
=\mathbb{E}\bigg[\lambda_1\Big(\mathcal{H}(\mathcal{P}_{\lambda_1},\,\mu)-\mathcal{H}(\mathcal{P}_{\lambda_1},\,\mu(\cdot|s))\Big)
\end{equation}
$$
 + \,\dots\,+\lambda_M\Big(\mathcal{H}(\mathcal{P}_{\lambda_M},\,\mu(\cdot|\cap_{i=1}^{M-1}\mathcal{P}_{\lambda_i}(\omega)))-\mathcal{H}(\mathcal{P}_{\lambda_M},\,\mu(\cdot|\cap_{i=1}^{M-1}\mathcal{P}_{\lambda_i}(\omega),\,s))\Big)     \bigg]
$$
\hypertarget{12}{}\begin{equation}
=\mathbb{E}\bigg[ \lambda_1 \Big( \mathcal{H}(\mathcal{P}_s,\,F(s)) - \mathcal{H}(\mathcal{P}_s,\,F(s|\mathcal{P}_{\lambda_1}(\omega)))\Big)
\end{equation}
$$
 +\,\dots\,+ \lambda_M\Big( \mathcal{H}(\mathcal{P}_s,\,F(s|\cap_{i=1}^{M-1}\mathcal{P}_{\lambda_i}(\omega))) - \mathcal{H}(\mathcal{P}_s,\,F(s|\cap_{i=1}^{M}\mathcal{P}_{\lambda_i}(\omega)))\Big)
\bigg]
$$

$$=\mathbb{E} \Big[ \lambda_1 \mathcal{H}(\mathcal{P}_s,\,F(s)) + (\lambda_2-\lambda_1)\mathcal{H}(\mathcal{P}_s,\,F(s|\mathcal{P}_{\lambda_1}(\omega)))$$
$$+\,\dots\,+ (\lambda_M-\lambda_{M-1})\mathcal{H}(\mathcal{P}_s,\,F(s|\cap_{i=1}^{M-1}\mathcal{P}_{\lambda_i}(\omega))) - \lambda_M\mathcal{H}(\mathcal{P}_s,\,F(s|\cap_{i=1}^{M}\mathcal{P}_{\lambda_i}(\omega)))  \Big]
$$

$$=\sum\limits_{\omega\in\Omega} \mu(\omega)\sum\limits_{n\in \mathcal{N}} \Big( - \lambda_1 \text{Pr}(n) \log (\text{Pr}(n)) - (\lambda_2 - \lambda_1) \text{Pr}(n|\mathcal{P}_{\lambda_1}(\omega))\log(\text{Pr}(n|\mathcal{P}_{\lambda_1}(\omega)))$$
$$ - (\lambda_3 - \lambda_2) \text{Pr}(n|\mathcal{P}_{\lambda_1}(\omega)\cap\mathcal{P}_{\lambda_2}(\omega))\log(\text{Pr}(n|\mathcal{P}_{\lambda_1}(\omega)\cap \mathcal{P}_{\lambda_2}(\omega))) 
$$
$$-\,\dots\,- (\lambda_M - \lambda_{M-1})\text{Pr}(n|\cap_{i=1}^{M-1} \mathcal{P}_{\lambda_i}(\omega))\log(\text{Pr}(n|\cap_{i=1}^{M-1} \mathcal{P}_{\lambda_i}(\omega)))
+ \lambda_M \text{Pr}(n|\omega)\log(\text{Pr}(n|\omega))\Big).
$$
\noindent The equality of (\hyperlink{11}{11}) and (\hyperlink{12}{12}) follows from the symmetry of mutual information, defined in \hyperlink{ap1}{Appendix 1}. $\blacksquare$
\bigskip

We now resume our proof of \hyperlink{le3}{Lemma 3}. First notice that \hyperlink{le9}{Lemma 9} establishes $\textbf{C}(\mathbb{P},\, \mu)$. For each $n\in \mathcal{N}$, let $s_n$ denote a signal in $S_n$ which results in the posterior generated by signals in $S_n$ with probability one (in \hyperlink{le8}{Lemma 8} we showed we can do this). Then notice:
$$\sum\limits_{\omega\in\Omega} \int\limits_s V(s)F(ds|\omega)\mu(\omega) = \sum\limits_{n\in\mathcal{N}} V(s_n)\int\limits_{s\in S_n} \sum\limits_{\omega\in\Omega} F(ds|\omega) \mu(\omega)
$$

$$=\sum\limits_{n\in\mathcal{N}} V(s_n)\text{Pr}(n) = \sum\limits_{n\in\mathcal{N}}\sum\limits_{\omega\in\Omega} \textbf{v}_n(\omega)F(\omega|s_n) \text{Pr}(n)
$$

$$= \sum\limits_{n\in\mathcal{N}} \sum\limits_{\omega\in\Omega} \mathbf{v}_n(\omega) \text{Pr}(n|\omega) \mu(\omega)
$$
\noindent Where the last step follows from the fact that $\text{Pr}(X|Y)\text{Pr}(Y)=\text{Pr}(Y|X)\text{Pr}(X)$. We now proceed with two proofs by contradiction. First, assume that $(F,\,a)$ is a solution to (\hyperlink{2}{2}) subject to (\hyperlink{3}{3}), which achieves expected utility $U_1$, and let $\mathbb{P}$ be the choice probabilities induced by it. Assume that $\mathbb{P}$ is not a solution to (\hyperlink{7}{7}) subject to (\hyperlink{8}{8}) and (\hyperlink{9}{9}), and thus there is a $\tilde{\mathbb{P}}$ which satisfies (\hyperlink{8}{8}) and (\hyperlink{9}{9}) and achieves expected utility $U_2> U_1$. However, a strategy pairing $(\tilde{F},\,\tilde{a})$ can be created that generates $\tilde{\mathbb{P}}$. For instance, for each of $N$ distinct signals ${s_n}$, let $\tilde{a}(\tilde{F}(\omega|{s_n}))\equiv n$, and let $\tilde{F}(s_n,\,\omega)=\tilde{\text{Pr}}(n|\omega) \mu(\omega)\,\,\,\forall\, \omega$ so that (\hyperlink{3}{3}) is satisfied. This is impossible though as then $(\tilde{F},\,\tilde{a})$ achieves $U_2> U_1$ and $(F,\,a)$ cannot have been optimal.

Similarly, assume that $\mathbb{P}$ is a solution to (\hyperlink{7}{7}) subject to (\hyperlink{8}{8}) and (\hyperlink{9}{9}), which achieves expected utility $U_3$ and but is not induced by a solution to \hyperlink{2}{2} subject to (\hyperlink{3}{3}). That is there is a $\tilde{F}$ which satisfies (\hyperlink{3}{3}) and achieves $U_4>U_3$. This means, however, that $\tilde{\text{Pr}}(n|\omega)=\dfrac{\tilde{F}(s_n,\,\omega)}{\mu(\omega)}$ also achieves $U_4$, which is impossible as $\mathbb{P}$ was supposedly optimal and $\tilde{\mathbb{P}}$ satisfies (\hyperlink{8}{8}) and (\hyperlink{9}{9}). $\blacksquare$

\hypertarget{pt2}{ }
 
\noindent \textbf{Proof of \hyperlink{th2}{Theorem 2}.} The Lagrangian for the above problem can be written:
$$\mathcal{L} = \sum\limits_{n\in\mathcal{N}} \sum\limits_{\omega\in\Omega} \mathbf{v}_n(\omega) \text{Pr}(n|\omega) \mu(\omega) - \textbf{C}(\mathbb{P},\, \mu) + \sum\limits_{n\in\mathcal{N}} \sum\limits_{\omega\in\Omega} \xi_n(\omega)\text{Pr}(n|\omega)\mu(\omega) 
$$
$$- \sum\limits_{\omega\in\Omega} \gamma(\omega)\Big(\sum\limits_{n\in\mathcal{N}} \text{Pr}(n|\omega) - 1 \Big)\mu(\omega)
$$
\noindent Where $\xi_n(\omega)\geq 0$ are the Lagrange multipliers for (\hyperlink{8}{8}), and $\gamma(\omega)$ are the multipliers for (\hyperlink{9}{9}). If $\text{Pr}(n)=0$, then $\text{Pr}(n|\omega)=0\,\,\forall\omega\in\Omega$. If $\text{Pr}(n|\cap_{i=1}^{m}\mathcal{P}_{\lambda_i}(\omega)) = 0$ for some $m\in\{1,\,\dots,\,M-1\}$ and $\omega$, then $\text{Pr}(n|\omega)=0$. If $\text{Pr}(n)>0,$ and $\text{Pr}(n|\cap_{i=1}^{m}\mathcal{P}_{\lambda_i}(\omega)) > 0,\,\forall m\in\{1,\,\dots,\,M-1\}$, then the first order condition with respect to $\text{Pr}(n|\omega)$ implies:
$$ \mathbf{v}_n(\omega) + \lambda_1(1+\log \text{Pr}(n)) + (\lambda_2 - \lambda_1)(1+\log \text{Pr}(n|\mathcal{P}_{\lambda_1}(\omega)))
$$
$$+\,\dots\,+(\lambda_M - \lambda_{M-1})(1+\log \text{Pr}(n|\cap_{i=1}^{M-1}\mathcal{P}_{\lambda_i}(\omega))) - \lambda_M (1+\log \text{Pr}(n|\omega)) = \gamma(\omega) - \xi_n(\omega)
$$
which then implies $\text{Pr}(n|\omega)>0$ and $\xi_n(\omega) =0$, because if not, and $\text{Pr}(n|\omega)=0$, then since $\xi_n(\omega)\geq0$, equality of the first order condition then necessitates $\gamma(\omega) = \infty $. This is impossible, however, since then $\forall \, \nu \in \mathcal{N}$ their respective first order conditions holding necessitates $ \text{Pr}(\nu|\omega) =0$. This being true $\forall \, \nu \in \mathcal{N}$ of course then violates (\hyperlink{9}{9}). Thus, the first order condition implies:
\hypertarget{13}{}\begin{equation}
\text{Pr}(n|\omega)= \text{Pr}(n)^{\frac{\lambda_1}{\lambda_M}}\text{Pr}(n|\mathcal{P}_{\lambda_1}(\omega))^\frac{\lambda_2 - \lambda_1}{\lambda_M}\dots\,\text{Pr}(n|\cap_{i=1}^{M-1}\mathcal{P}_{\lambda_i}(\omega))^\frac{\lambda_M - \lambda_{M-1}}{\lambda_M} \mathlarger{\mathlarger{e}} ^\frac{\mathbf{v}_n(\omega)}{\lambda_M}  \mathlarger{\mathlarger{e}} ^\frac{-\gamma(\omega)}{\lambda_M}
\end{equation}
\noindent Plugging (\hyperlink{13}{13}) into (\hyperlink{9}{9}), one can solve for $\gamma(\omega)$. Plugging $\gamma(\omega)$ back into (\hyperlink{13}{13}) achieves the desired result.$\blacksquare$

\bigskip

\noindent \textbf{Proof of \hyperlink{co1}{Corollary 1}.} Plug equation (\hyperlink{10}{10}) into equation (\hyperlink{7}{7}). $\blacksquare$

\bigskip
 
 \hypertarget{pt3}{ }
\noindent \textbf{Proof of \hyperlink{th3}{Theorem 3}.} A fixed effect interpretation of MSSE follows easily from the optimal choice probabilities described in \hyperlink{th2}{Theorem 2}:
$$\text{Pr}(n|\omega) = \dfrac{\text{Pr}(n)^{\frac{\lambda_1}{\lambda_M}}\text{Pr}(n|\mathcal{P}_{\lambda_1}(\omega) )^\frac{\lambda_2 - \lambda_1}{\lambda_M}\dots\,\text{Pr}(n|\cap_{i=1}^{M-1}\mathcal{P}_{\lambda_i}(\omega))^\frac{\lambda_M - \lambda_{M-1}}{\lambda_M} \mathlarger{\mathlarger{e}} ^\frac{\mathbf{v}_n(\omega)}{\lambda_M}}{\mathlarger{\sum}\limits_{\nu\in\mathcal{N}}    \text{Pr}(\nu)^{\frac{\lambda_1}{\lambda_M}}\text{Pr}(\nu|\mathcal{P}_{\lambda_1}(\omega) )^\frac{\lambda_2 - \lambda_1}{\lambda_M}\dots\,\text{Pr}(\nu|\cap_{i=1}^{M-1}\mathcal{P}_{\lambda_i}(\omega))^\frac{\lambda_M - \lambda_{N-1}}{\lambda_M} \mathlarger{\mathlarger{e}} ^\frac{\mathbf{v}_\nu(\omega)}{\lambda_M}}
$$
$$=\dfrac{(N\text{Pr}(n))^{\frac{\lambda_1}{\lambda_M}}(N\text{Pr}(n|\mathcal{P}_{\lambda_1}(\omega) ))^\frac{\lambda_2 - \lambda_1}{\lambda_M}\dots\,(N\text{Pr}(n|\cap_{i=1}^{M-1}\mathcal{P}_{\lambda_i}(\omega)))^\frac{\lambda_M - \lambda_{M-1}}{\lambda_M} \mathlarger{\mathlarger{e}} ^\frac{\mathbf{v}_n(\omega)}{\lambda_M}}{\mathlarger{\sum}\limits_{\nu\in\mathcal{N}}   (N\text{Pr}(\nu))^{\frac{\lambda_1}{\lambda_M}}(N\text{Pr}(\nu|\mathcal{P}_{\lambda_1}(\omega) ))^\frac{\lambda_2 - \lambda_1}{\lambda_M}\dots\,(N\text{Pr}(\nu|\cap_{i=1}^{M-1}\mathcal{P}_{\lambda_i}(\omega)))^\frac{\lambda_M - \lambda_{N-1}}{\lambda_M} \mathlarger{\mathlarger{e}} ^\frac{\mathbf{v}_\nu(\omega)}{\lambda_M}}
$$
$$=\mathlarger{\dfrac{\mathlarger{\mathlarger{e} ^\frac{\mathbf{v}_n(\omega)+ \lambda_1 \alpha_n^0 + (\lambda_2-\lambda_1)\alpha_n^1 +\dots+(\lambda_M - \lambda_{M-1})\alpha_n^{M-1}}{\lambda_M}}}{\mathlarger{\sum}\limits_{\nu\in\mathcal{N}} \mathlarger{\mathlarger{e} ^\frac{\mathbf{v}_\nu(\omega)+ \lambda_1 \alpha_\nu^0 + (\lambda_2-\lambda_1)\alpha_\nu^1 +\dots+(\lambda_M - \lambda_{M-1})\alpha_\nu^{M-1}}{\lambda_M}}}}
$$
\noindent Where $\alpha_\nu^0 = \log (N \text{Pr}(\nu))$, and for $m\in \{1,\,\dots,\,M-1\}$ we have $\alpha_\nu^m = \log( N \text{Pr}(\nu|\cap_{i=1}^m \mathcal{P}_{\lambda_i}(\omega)))$. Normalizing the value of the options by $\lambda_M$, namely letting $\tilde{v}_n= \frac{\mathbf{v}_n(\omega)}{\lambda_M}$, and defining $\alpha_n$ appropriately, agent choice behavior described by rational inattention with MSSE can then be interpreted as a RU model where each option $n$ has perceived value:
$$u_n = \tilde{v}_n + \mathsmaller{\mathsmaller{\frac{\lambda_1}{\lambda_M}}} \alpha_n^0 + \mathsmaller{\mathsmaller{\frac{\lambda_2-\lambda_1}{\lambda_M}}}\alpha_n^1+\dots+ \mathsmaller{\mathsmaller{\frac{\lambda_M-\lambda_{M-1}}{\lambda_M}}}\alpha_n^{M-1} +\epsilon_n = \tilde{v}_n + \alpha_n + \epsilon_n
$$
\noindent The only kind of RU model consistent with this behavior is one where $\epsilon_n$ is distributed iid according to a Gumbel distribution \cite{tra09}. $\blacksquare$

\hypertarget{ap3}{ }
\section*{Appendix 3}

The behavior described in \hyperlink{th2}{Theorem 2} has many intuitive features. It is also a quite natural extension of the analogous result from \citeA{matmck15}, which is described in equation (\hyperlink{14}{14}). If we assume the agent has prior $\mu$, and all partitions are learning strategy invariant (the environment studied in \citeA{matmck15}) and have associated multiplier $\lambda_2$, then if the agent does optimal research in state $\omega\in\Omega$, they select option $n$ from their set of options $\mathcal{N}$ with probability:
\hypertarget{14}{}\begin{equation}
\text{Pr}(n|\omega) = \dfrac{\text{Pr}(n) \mathlarger{\mathlarger{e}} ^\frac{\mathbf{v}_n(\omega)}{\lambda_2}}{\sum\limits_{\nu\in\mathcal{N}}\text{Pr}(\nu) \mathlarger{\mathlarger{e}} ^\frac{\mathbf{v}_\nu(\omega)}{\lambda_2}}.
\end{equation}

One major takeaway from the formula in (\hyperlink{14}{14}) is that when Shannon Entropy is used to measure uncertainty the chance of the agent selecting an option $n$ in a particular state of the world $\omega$ is fully determined by the unconditional chances of the options being selected, $\text{Pr}(n)$, and the realized values of the options in that state of the world. Beyond this takeaway, the formula in (\hyperlink{14}{14}) also has many intuitive features. If $\lambda_2$ grows, which represents an increase in the difficulty of learning, the value of each option in the realized state becomes less significant for the determination of the selected option, and the significance of the agent's prior increases. Similarly, if $\lambda_2$ shrinks, the agent's prior becomes less significant, and the realized values of the options becomes more significant. If $\lambda_2$ approaches infinity, the realized values become insignificant, and the behavior of the agent approaches the behavior of the agent in the case where learning is impossible: they choose their option based on their prior. If $\lambda_2$ approaches zero the unconditional priors become insignificant, and the behavior of the agent approaches the behavior of the agent in the case where learning is costless: they choose the option with the highest realized value.

If we instead assume that the agent may also learn through a partition with a lower multiplier $\lambda_1$, that can convey information about the realization $\mathcal{P}_{\lambda_1}(\omega)$ of a partition $\mathcal{P}_{\lambda_1}$ of $\Omega$, then if $\mathcal{P}_{\lambda_1}\neq \Omega$, and the agent does optimal research in state $\omega\in\Omega$, they select option $n$ from their set of options $\mathcal{N}$ with probability:
\hypertarget{15}{}
\begin{equation}
\text{Pr}(n|\omega) = \dfrac{\text{Pr}(n)^{\frac{\lambda_1}{\lambda_2}} \text{Pr}(n|\mathcal{P}_{\lambda_1}(\omega))^{\frac{\lambda_2-\lambda_1}{\lambda_2}} \mathlarger{\mathlarger{e}} ^\frac{\mathbf{v}_n(\omega)}{\lambda_2}}{\sum\limits_{\nu\in\mathcal{N}}\text{Pr}(\nu)^{\frac{\lambda_1}{\lambda_2}} \text{Pr}(\nu|\mathcal{P}_{\lambda_1}(\omega))^{\frac{\lambda_2-\lambda_1}{\lambda_2}} \mathlarger{\mathlarger{e}} ^\frac{\mathbf{v}_\nu(\omega)}{\lambda_2}}.
\end{equation}

With MSSE, as the formula in (\hyperlink{15}{15}) indicates, the chance of the agent selecting an option $n$ in a particular state of the world $\omega$ depends not only on the unconditional chances of the options being selected and the realized values of the options, but also on the values that the options take in similar states of the world, states that result in the same realization of $\mathcal{P}_{\lambda_1}$. When option $n$ is in general desirable in $\mathcal{P}_{\lambda_1}(\omega)$ relative to the other options, then $\text{Pr}(n|\mathcal{P}_{\lambda_1}(\omega))$ is larger, and there may be a high chance of $n$ being selected, even if $\text{Pr}(n)$ is not that large, and $\textbf{v}_n(\omega)$ is not that high.

The formula in (\hyperlink{15}{15}) also has many intuitive features. It maintains the intuitive comparative statistics for $\lambda_2$ that the formula in (\hyperlink{14}{14}) had, and also features intuitive properties for $\text{Pr}(n|\mathcal{P}_{\lambda_1}(\omega))$ and $\lambda_1$. If observing $\mathcal{P}_{\lambda_1}(\omega)$ is completely uninformative about the value of the options, then it is optimal for the agent to select $\text{Pr}(n|\mathcal{P}_{\lambda_1}(\omega))=\text{Pr}(n)$ since $\min\limits_{S\in S(\Omega)}C(S,\,\mu)$ is strictly concave in $\mu$. In this case $\text{Pr}(n)^{\frac{\lambda_1}{\lambda_2}} \text{Pr}(n|\mathcal{P}_{\lambda_1}(\omega))^{\frac{\lambda_2-\lambda_1}{\lambda_2}}=\text{Pr}(n),$ and behavior is identical to that in (\hyperlink{14}{14}). If the cheaper information source contains irrelevant information it is thus ignored, and behavior collapses back to the environment described in \citeA{matmck15}, as we should desire. If $\lambda_1$ approaches $\lambda_2$ (the cheaper information source becomes close to as expensive as the more expensive information source) then behavior approaches that described in (\hyperlink{14}{14}) since $\text{Pr}(n)^{\frac{\lambda_1}{\lambda_2}} \text{Pr}(n|\mathcal{P}_{\lambda_1}(\omega))^{\frac{\lambda_2-\lambda_1}{\lambda_2}}\rightarrow\text{Pr}(n)$. Thus, if an insignificantly cheaper information source is introduced behavior is changed in an insignificant fashion. Again, this seems like a desirable property. If $\lambda_1$ approaches zero then the role of the unconditional priors dissipates, and exponent on $\text{Pr}(n|\mathcal{P}(\omega))$ approaches one, meaning it replaces the unconditional prior from (\hyperlink{14}{14}). This makes sense because if $\lambda_1$ goes to zero it means $\mathcal{P}_{\lambda_1}(\omega)$ can essentially be viewed for free, in which case behavior within each $\mathcal{P}_{\lambda_1}(\omega)$ should resemble that in the setting where there is only one information source with multiplier $\lambda_2$ and a prior of $\mu(\cdot|\mathcal{P}_{\lambda_1}(\omega))$.

We can continue adding as many new partitions with new associated multipliers as we desire and the description of behavior in \hyperlink{th2}{Theorem 2} maintains the sorts of intuitive properties described in the paragraphs above. RI with MSSE is thus a very natural extension of RI with Shannon Entropy.

\newpage

\bibliographystyle{apacite}

\bibliography{ref}

\end{document}